\documentclass[12pt]{article}
\usepackage{graphicx}
\usepackage{here}
\newcommand{\lesssim}{\raisebox{0.3mm}{\em $\, <$} \hspace{-3.2mm}
\raisebox{-1.4mm}{\em $\sim \,$}}
\newcommand{\gtrsim}{\raisebox{0.3mm}{\em $\, >$} \hspace{-3.2mm}
\raisebox{-1.4mm}{\em $\sim \,$}}
\begin{document}

\begin{center}
{\large\bf
  Neutrino Oscillations at low energy long baseline experiments
  in the presence of nonstandard interactions and parameter degeneracy}
\end{center}
\vspace*{.6cm}

\begin{center}
\large{\sc Osamu Yasuda}
\end{center}
\vspace*{0cm}
{\it
\begin{center}
Department of Physics, Tokyo Metropolitan University,\\
  Hachioji, Tokyo 192-0397, Japan
\end{center}}

\vspace*{0.5cm}

{\Large \bf
\begin{center} Abstract \end{center}  }

  We discuss the analytical expression
  of the oscillation probabilities at low energy long baseline experiments,
  such as T2HK and T2HKK in the presence of
  nonstandard interactions (NSIs).  We show that these experiments are
  advantageous to explore the NSI parameters
  ($\epsilon_D$, $\epsilon_N$), which were suggested to be nonvanishing
  to account for the discrepancy between the solar neutrino and KamLAND data.
  We also show that, when the NSI parameters are small,
  parameter degeneracy in the CP phase $\delta$,
  $\epsilon_D$ and $\epsilon_N$ can be resolved by combining data of
  the T2HK and T2HKK experiments.

\vspace*{.5cm}
\newpage
\section{Introduction}
In the last two decades
we have been successful in determination of the oscillation
parameters in the standard
three flavor framework\,\cite{Tanabashi:2018oca}.
The three flavor neutrino oscillation is described by
the mixing matrix
\begin{eqnarray}
U=\left(
\begin{array}{ccc}
c_{12}c_{13} & s_{12}c_{13} &  s_{13}e^{-i\delta}\nonumber\\
-s_{12}c_{23}-c_{12}s_{23}s_{13}e^{i\delta} & 
c_{12}c_{23}-s_{12}s_{23}s_{13}e^{i\delta} & s_{23}c_{13}\nonumber\\
s_{12}s_{23}-c_{12}c_{23}s_{13}e^{i\delta} & 
-c_{12}s_{23}-s_{12}c_{23}s_{13}e^{i\delta} & c_{23}c_{13}\nonumber\\
\end{array}\right),
\nonumber
\end{eqnarray}
where the following notations are adopted:
$c_{jk}\equiv\cos\theta_{jk}$,
$s_{jk}\equiv\sin\theta_{jk}$ and 
$\theta_{jk}$~$((j,k)=(1,2), (1,3), (2,3))$
are the three mixing angles and $\delta$ is
the CP phase.
The mixing angles $\theta_{12}$, $\theta_{13}$
and the two mass squared differences
$\Delta m^2_{21}$, $|\Delta m^2_{31}|$
have been measured with good
precision\,\cite{Capozzi:2018ubv,Esteban:2018azc,Valle:2018pgs},
while the uncertainty in $\theta_{23}$ and $\delta$
is still large.
Furthermore, the mass hierarchy
(whether the mass pattern is given by
normal hierarchy or inverted hierarchy)
and the octant of $\theta_{23}$ (whether $\theta_{23}$ is
larger than $\pi/4$ or not) is not known,
although the normal hierarchy and the higher octant
$\theta_{23} > \pi/4$ are favored to some
extent\,\cite{Capozzi:2018ubv,Esteban:2018azc,Valle:2018pgs}.
The uncertainties in these oscillation parameters
are expected to be much reduced in the future long baseline
experiments, T2HK\,\cite{Abe:2014oxa} at $L$=295km,
T2HKK\,\cite{Abe:2016ero} at $L$=1100km and
DUNE\,\cite{Acciarri:2015uup} at $L$=1300km.

On the other hand,
there have been a few experimental results
which do not seem to be explained by
the standard three flavor framework.
One of them is the tension between the mass squared difference
from the solar neutrino experiments and
the KamLAND data.  It has been pointed out
that this tension can be removed by introducing
either a nonstandard interaction (NSI)
in the neutrino propagation\,\cite{Gonzalez-Garcia:2013usa,Esteban:2018ppq} or
sterile neutrinos with mass squared difference
of O($10^{-5}$) eV$^2$\,\cite{Maltoni:2015kca}.\,\footnote{
  See Refs.\,\cite{Ohlsson:2012kf,Miranda:2015dra,Dev:2019anc}
  on NSI and Ref.\,\cite{Abazajian:2012ys}
  on sterile neutrino
  for extensive references.}

To know whether Nature is described by the NSI scenario discussed
in Ref.\,\cite{Gonzalez-Garcia:2013usa},
it is important to investigate how to check it.
In the analysis of the long-baseline experiments and
the atmospheric neutrino experiments, the dominant
oscillation comes from the larger mass squared difference
$\Delta m_{31}^2$ and the oscillation probabilities are
expressed in terms of $\epsilon_{\alpha\beta}$, which
will be defined in Eq.\,(\ref{eps1}) below, in addition to
the standard oscillation parameters.
While the results in Ref.~\cite{Gonzalez-Garcia:2013usa}
may suggest the existence of the NSI,
the parametrization for the NSI parameters
($\epsilon_D$, $\epsilon_N$), which
will be defined in Eq.\,(\ref{epsdn}) below,
is different
from the one with $\epsilon_{\alpha\beta}$ and
it is not clear how the allowed region in Ref.~\cite{Gonzalez-Garcia:2013usa}
will be tested or excluded by the future experiments.
In the past there were a couple of attempts to
estimate the sensitivity of the future experiments
to ($\epsilon_D$, $\epsilon_N$).
In Ref.\,\cite{Fukasawa:2016nwn}, assuming the standard oscillation scenario,
the excluded region in the ($\epsilon_D$, $\epsilon_N$)-plane
by the atmospheric neutrino measurements at Hyper-Kamiokande was given.
Ref.\,\cite{Ghosh:2017lim} estimated the sensitivity of future long
baseline experiments in testing the current best fit point suggested
by solar neutrino data.

In this paper we discuss the analytical expression
of the oscillation probabilities in the presence of the NSI
at low energy neutrino measurements ($\lesssim$ 1GeV),
such as T2HK and T2HKK, and show that low energy
neutrino measurements are advantageous because the
oscillation probabilities involve the fewer NSI
parameters including $\epsilon_D$, $\epsilon_N$.
The oscillation probabilities at low energy
in the presence of the NSI was discussed in Ref.\,\cite{Ge:2016dlx}
from a different point of view.
The oscillation probabilities at higher energy
experiments, such as DUNE, involve more NSI parameters
and discussions at higher energy are left as a future work.
We also show how parameter degeneracy can be
resolved by combining data at different
baseline length and different energy in the
T2HK and T2HKK system.
Parameter degeneracy in the presence of
the NSI is a complicated problem
and has been discussed by many
people\,\cite{Gago:2009ij,Coloma:2011rq,Bakhti:2014pva,Mocioiu:2014gua,Coloma:2015kiu,Liao:2016hsa,Blennow:2016etl,Agarwalla:2016fkh,Ge:2016dlx,Deepthi:2016erc,Liao:2016orc,Masud:2018pig,Verma:2019oeb}.
The situation of parameter degeneracy in low energy long baseline
experiments is better than that at high energy, because the
oscillation probabilities at low energy involve fewer numbers of the
NSI parameters.

\section{Nonstandard interactions in propagation}
Suppose that we have a 
flavor-dependent neutral current
NSI\,\cite{Wolfenstein:1977ue,Valle:1987gv,Guzzo:1991hi,Roulet:1991sm}:
\begin{eqnarray}
  &{\ }&\hspace*{-60mm}
{\cal L}_{\mbox{\tiny{\rm NSI}}} 
=-2\sqrt{2}\, \epsilon_{\alpha\beta}^{ff'P} G_F
\left(\overline{\nu}_{\alpha L} \gamma_\mu \nu_{\beta L}\right)\,
\left(\overline{f}_P \gamma^\mu f_P'\right)\,,
\nonumber
\end{eqnarray}
where $f_P$ and $f_P'$ are the fermions with chirality
$P=(1\pm\gamma_5)/2$,
$\epsilon_{\alpha\beta}^{ff'P}$ is a dimensionless constant
normalized in terms of the Fermi coupling constant
$G_F$.  Then, the matter potential in the flavor basis
is modified as
\begin{eqnarray}
  &{\ }&\hspace*{-50mm}
  {\cal A} =
A\left(
\begin{array}{ccc}
  1&0&0\\
  0&0&0\\
  0&0&0
\end{array}
\right)+  A\sum_{f=e,u,d}\frac{N_f}{N_e}\left(
\begin{array}{ccc}
\epsilon^f_{ee} & \epsilon^f_{e\mu} & \epsilon^f_{e\tau}\\
\epsilon^f_{\mu e} & \epsilon^f_{\mu\mu} & \epsilon^f_{\mu\tau}\\
\epsilon^f_{\tau e} & \epsilon^f_{\tau\mu} & \epsilon^f_{\tau\tau}
\end{array}
\right),
\label{matter-np-solar}
\end{eqnarray}
where $A\equiv\sqrt{2}G_F N_e$,
the new NSI parameters is defined as
$\epsilon_{\alpha\beta}^{f}\equiv
\epsilon_{\alpha\beta}^{ffL}+\epsilon_{\alpha\beta}^{ffR}
$,
since the matter effect is sensitive only to the coherent scattering,
and only to the vector part in the interaction, and
$N_f$ stands for number density of fermion $f$,
where $f$ is assumed to be u or d quarks or electrons.

In the case of solar neutrino
analysis\,\cite{Gonzalez-Garcia:2013usa,Esteban:2018ppq},
since the ratio of the density of protons to that of neutrons
varies along the neutrino path, 
the case with $\epsilon_{\alpha\beta}^{u}\ne0$, the one with
$\epsilon_{\alpha\beta}^{d}\ne0$, or the one
with both must be analyzed separately.\footnote{
The case with $\epsilon_{\alpha\beta}^{e}\ne 0$ was not considered
in Refs.\,\cite{Gonzalez-Garcia:2013usa,Esteban:2018ppq}
because of the complication in which
the NSI $\epsilon_{\alpha\beta}^{e}$ would also affect the
rate of the interactions between neutrinos and electrons
at detection.}
On the other hand, in the case of
atmospheric neutrinos or accelerator-based
long baseline neutrinos, which go through
the Earth, we can assume approximately that
the numbers of density for electrons, protons
and neutrons are almost equal,
$N_e\simeq N_p\simeq N_n$.
So in this case, the matter
potential (\ref{matter-np-solar})
can be written as
\begin{eqnarray}
  &{\ }&\hspace*{-60mm}
{\cal A} = A\left(
\begin{array}{ccc}
1+ \epsilon_{ee} & \epsilon_{e\mu} & \epsilon_{e\tau}\\
\epsilon_{\mu e} & \epsilon_{\mu\mu} & \epsilon_{\mu\tau}\\
\epsilon_{\tau e} & \epsilon_{\tau\mu} & \epsilon_{\tau\tau}
\end{array}
\right),
\label{matter-np}
\end{eqnarray}
where the new parameter $\epsilon_{\alpha\beta}$ is defined as
\begin{eqnarray}
  &{\ }&\hspace*{-40mm}
  \epsilon_{\alpha\beta}\,\equiv\,
  \sum_{f=e,u,d}
\frac{N_f}{N_e}\epsilon_{\alpha\beta}^{f}
\simeq \epsilon_{\alpha\beta}^e+3\epsilon_{\alpha\beta}^u
+3\epsilon_{\alpha\beta}^d\,.
\label{eps1}
\end{eqnarray}

While the constraints on $\epsilon_{\alpha\beta}^{f}$
by various experiments
except neutrino oscillations were
given in Refs.\,\cite{Davidson:2003ha,Biggio:2009nt},
the updated bounds on $\epsilon_{\alpha\beta}$
by global analysis of oscillation experiments
are given in Ref.\,\cite{Esteban:2018ppq}.
The allowed region for $\epsilon_{\alpha\beta}$
at 90\% CL can be read off
from Fig.\,9 in Ref.\,\cite{Esteban:2018ppq} as
follows:
\begin{eqnarray}
  &{\ }&\hspace*{-30mm}
\left\{\begin{array}{rcl}
  -0.21< & \epsilon_{ee}-\epsilon_{\mu\mu}
  &<0.26
  \\
  -0.018<&\epsilon_{\tau\tau}-\epsilon_{\mu\mu}
  &<0.071
  \\
  -0.10<&\epsilon_{e\mu}
  &<0.10
  \\
-0.25<&\epsilon_{e\tau}
  &<0.063
  \\
-0.015<&\epsilon_{\mu\tau}
&<0.021
\end{array}\right\}\qquad
(90\%\mbox{\rm CL})
  \label{eab}
\end{eqnarray}  

\section{Oscillation probabilities at low energy}

\subsection{Solar neutrino flavor basis}
At low energy $E\,\lesssim$1GeV, the condition
\begin{eqnarray}
  &{\ }&\hspace*{-70mm}
  \frac{\Delta m^2_{21}}{2E}\sim A \,\ll
  \frac{|\Delta m^2_{31}|}{2E}\,,
  \nonumber
\end{eqnarray}
is satisfied and the ratio of the two scales
is approximately given by
$\Delta m^2_{21}/|\Delta m^2_{31}|\simeq 1/30$.
So the oscillation probability can be
expressed analytically by a perturbation method
with respect to this ratio.

At low energy it is convenient\,\cite{Ge:2016dlx}
to change the flavor basis into the solar neutrino flavor basis.
The $3 \times 3$ Hamiltonian can be written as
\begin{eqnarray}
  &{\ }&\hspace*{-30mm}
  H=R_{23}\, \tilde{R}_{13}
  \,R_{12}\,\mbox{\rm diag}
  \left(0,\frac{\Delta m^2_{21}}{2E},\frac{\Delta m^2_{31}}{2E}\right)
  \,R_{12}^{-1}\,\tilde{R}_{13}^{-1}
  \,R_{23}^{-1}+{\cal A}
  \nonumber\\
  &{\ }&\hspace*{-25mm}
  =R_{23}\, \tilde{R}_{13}\,\left[R_{12}\,\mbox{\rm diag}
  \left(0,\frac{\Delta m^2_{21}}{2E},\frac{\Delta m^2_{31}}{2E}\right)
  \,R_{12}^{-1}
  +{\cal A}'\right]\,\tilde{R}_{13}^{-1}\,R_{23}^{-1}\,,
  \label{h1}
\end{eqnarray}
where
\begin{eqnarray}
  &{\ }&\hspace*{-25mm}
R_{23}\equiv\exp(i\theta_{23}\lambda_7)\,,
\nonumber\\
  &{\ }&\hspace*{-25mm}
\tilde{R}_{13}\equiv\mbox{\rm diag}(e^{-i\delta/2},1,e^{i\delta/2})\,
\exp(i\theta_{13}\lambda_5)\,
\mbox{\rm diag}(e^{i\delta/2},1,e^{-i\delta/2})\,,
\nonumber\\
  &{\ }&\hspace*{-25mm}
R_{12}\equiv\exp(i\theta_{12}\lambda_2)
\nonumber
\end{eqnarray}
are the $3\times 3$ rotational matrices,
\begin{eqnarray}
  &{\ }&\hspace*{-25mm}
\lambda_2\equiv\left(
\begin{array}{ccc}
0&-i&0\cr
i&0&0\cr
0&0&0
\end{array}\right),\quad
\lambda_5\equiv\left(
\begin{array}{ccc}
0&0&-i\cr
0&0&0\cr
i&0&0
\end{array}\right),\quad
\lambda_7\equiv\left(
\begin{array}{ccc}
0&0&0\cr
0&0&-i\cr
0&i&0
\end{array}\right)\nonumber
\end{eqnarray}
are the $3 \times 3$ Gell-Mann matrices,
and the matter potential ${\cal A}'$ in the solar neutrino
flavor basis is defined as
\begin{eqnarray}
  &{\ }&\hspace*{-25mm}
  {\cal A}'\equiv \tilde{R}_{13}^{-1}\,R_{23}^{-1}\,{\cal A}\,
  R_{23}\, \tilde{R}_{13}
  \nonumber\\
  &{\ }&\hspace*{-20mm}
  \equiv
A\left(
\begin{array}{ccc}
  c^2_{13}&0&e^{-i\delta}c_{13}s_{13}\\
  0&0&0\\
  e^{i\delta}c_{13}s_{13}&0&s^2_{13}
\end{array}
\right)+  A\sum_{f=e,u,d}\frac{N_f}{N_e}\left(
\begin{array}{ccc}
{\epsilon^f_{11}}' & {\epsilon^f_{12}}' & {\epsilon^f_{13}}'\\
{\epsilon^f_{21}}' & {\epsilon^f_{22}}' & {\epsilon^f_{23}}'\\
{\epsilon^f_{31}}' & {\epsilon^f_{23}}' & {\epsilon^f_{33}}'
\end{array}
\right)\,.
  \label{ap1}
\end{eqnarray}
Because solar neutrinos are approximately
driven by one mass squared difference,
the analysis of solar neutrinos with
the $3 \times 3$ Hamiltonian (\ref{h1})
is reduced to that of the following
effective $2 \times 2$ Hamiltonian\,\cite{Gonzalez-Garcia:2013usa}:
\begin{eqnarray*}
  &{\ }&\hspace*{-5mm}
H^{\rm eff}=
\frac{\Delta m^2_{21}}{4E}\left(\begin{array}{cc}
-\cos2\theta_{12} & \sin2\theta_{12}  \\
\sin2\theta_{12} & \cos2\theta_{12}
\end{array}\right) 
+
\left(\begin{array}{cc}
A c^2_{13} & 0 \\
0 & 0
\end{array}\right)  + 
 A\sum_{f=e,u,d}\frac{N_f}{N_e}
\left(\begin{array}{cc}
- \epsilon_D^f &  \epsilon_N^f \\
 \epsilon_N^{f*} &  \epsilon_D^f
\end{array}\right),
\end{eqnarray*}
where  $\epsilon^f_{D}$ and $\epsilon^f_{N}$ are
related to the components of ${\cal A}'$:
\begin{eqnarray}
&{\ }&\hspace*{-80mm}
  \epsilon_D^f =\frac{1}{2}\left(
  {\epsilon^{f}_{22}}'-{\epsilon^{f}_{11}}'\right)\,,\quad
\epsilon_N^f = {\epsilon^{f}_{12}}'\,.
\label{epsdn}
\end{eqnarray}
It has been pointed out that the value of $\Delta m^2_{21}$
inferred from the solar neutrino data and
that from the KamLAND experiment have a tension
at 2$\sigma$, and the results of
Refs.\,\cite{Gonzalez-Garcia:2013usa,Esteban:2018ppq}
show that a nonvanishing value of 
$(\epsilon_D^f, \epsilon_N^f)$ solves this tension.
This gives a motivation to take NSI in
propagation seriously.

\subsection{Oscillation probability in the Earth}
To discuss low energy neutrino oscillations in the Earth,
let us introduce the Hamiltonian for neutrinos ($H^{(-)})$
  and for antineutrinos ($H^{(+)})$ in the solar flavor basis:
\begin{eqnarray}
  &{\ }&\hspace*{-20mm}
  H^{(\mp)}=R_{23}\, \tilde{R}_{13}^{(\mp)}\,\left[R_{12}\,\mbox{\rm diag}
  \left(0,\frac{\Delta m^2_{21}}{2E},\frac{\Delta m^2_{31}}{2E}\right)
  \,R_{12}^{-1}
  \mp({\cal A}^{(\mp)})'\right]\,(\tilde{R}_{13}^{(\mp)})^{-1}\,R_{23}^{-1}\,,
    \label{h2}
\end{eqnarray}
where
\begin{eqnarray}
  &{\ }&\hspace*{-20mm}
\tilde{R}_{13}^{(\mp)}\equiv
\mbox{\rm diag}(e^{\mp i\delta/2},1,e^{\pm i\delta/2})\,
\exp(i\theta_{13}\lambda_5)\,
\mbox{\rm diag}(e^{\pm i\delta/2},1,e^{\mp i\delta/2})
  \nonumber\\
  &{\ }&\hspace*{-24mm}
  ({\cal A}^{(\mp)})'\equiv (\tilde{R}_{13}^{(\mp)})^{-1}\,R_{23}^{-1}\,{\cal A}\,
  R_{23}\, \tilde{R}_{13}^{(\mp)}
  \nonumber\\
  &{\ }&\hspace*{-10mm}
  = A\left(
\begin{array}{ccc}
  c^2_{13}&0&e^{\mp i\delta}c_{13}s_{13}\\
  0&0&0\\
  e^{\pm i\delta}c_{13}s_{13}&0&s^2_{13}
\end{array}
\right)+  A\left(
\begin{array}{ccc}
(\epsilon_{11}^{(\mp)})' & (\epsilon_{12}^{(\mp)})' & (\epsilon_{13}^{(\mp)})'\\
(\epsilon_{21}^{(\mp)})' & (\epsilon_{22}^{(\mp)})' & (\epsilon_{23}^{(\mp)})'\\
(\epsilon_{31}^{(\mp)})' & (\epsilon_{23}^{(\mp)})' & (\epsilon_{33}^{(\mp)})'
\end{array}
\right)\,,
  \label{ap2}
\end{eqnarray}
and we have defined the NSI parameters in the solar neutrino
basis for neutrinos $(\epsilon_{jk}^{(-)})'$ and antineutrinos
$(\epsilon_{jk}^{(+)})'$ separately:
\begin{eqnarray}
  &{\ }&\hspace*{-40mm}
  (\epsilon_{jk}^{(\mp)})'\,\equiv\,
  \sum_{f=e,u,d}
  \frac{N_f}{N_e}
\left[(\tilde{R}_{13}^{(\mp)})^{-1}\,R_{23}^{-1}\right]_{j\alpha}\,
  \epsilon_{\alpha\beta}^{f}
  \left[R_{23}\, \tilde{R}_{13}^{(\mp)}\right]_{\beta k}\,.
\label{eps2}
\end{eqnarray}
In practice, however,
the difference between $(\epsilon_{jk}^{(-)})'$
for neutrinos and $(\epsilon_{jk}^{(+)})'$
for antineutrinos is multiplied by a small
factor $s_{13}$, and because the constraints
(\ref{eab}) show that $\epsilon_{\alpha\beta}$ are
small, the difference between $(\epsilon_{jk}^{(-)})'$
for neutrinos and $(\epsilon_{jk}^{(+)})'$ is very small.
So we will identify $(\epsilon_{jk}^{(+)})'$ with
$(\epsilon_{jk}^{(-)})'$ and denote them simply
as $\epsilon_{jk}'$ in the following discussions
for simplicity.  Thus we have the Hamiltonian
for neutrinos and for antineutrinos in the solar flavor basis:
\begin{eqnarray}
  &{\ }&\hspace*{-20mm}
  H^{(\mp)}=R_{23}\, \tilde{R}_{13}^{(\mp)}\,\left[R_{12}\,\mbox{\rm diag}
  \left(0,\frac{\Delta m^2_{21}}{2E},\frac{\Delta m^2_{31}}{2E}\right)
  \,R_{12}^{-1}
  \mp{\cal A}'\right]\,(\tilde{R}_{13}^{(\mp)})^{-1}\,R_{23}^{-1}\,,
    \label{h3}
\end{eqnarray}
where ${\cal A}'$ is defined in Eq.\,(\ref{ap1}).

The oscillation probabilities are given by
(See Appendix\,\ref{appendixa} for details.)
\begin{eqnarray}
  &{\ }&\hspace{-6mm}
  \left\{\begin{array}{l}
P(\nu_{\mu}\to\nu_{e})\\
P(\bar{\nu}_{\mu}\to \bar{\nu}_e)
\end{array}\right\}
\nonumber\\
&{\ }&\hspace*{-10mm}
=4 \left|
\tilde{U}_{e3}^{(\mp)}\tilde{U}_{\mu 3}^{(\mp)\ast}
\sin\left(\frac{\Delta \tilde E_{31}^{(\mp)} L}{2}\right)
+e^{i\Delta \tilde E_{32}^{(\mp)} L/2}
\tilde{U}_{e2}^{(\mp)}\tilde{U}_{\mu 2}^{(\mp)\ast }
\sin\left(\frac{\Delta \tilde E_{21}^{(\mp)} L}{2}\right)
\right|^2
\label{pme}\\
&{\ }&\hspace*{-6mm}
\left\{\begin{array}{l}
P(\nu_{\mu}\to\nu_{\mu})\\
P(\bar{\nu}_{\mu}\to \bar{\nu}_\mu)
\end{array}\right\}
\nonumber\\
&{\ }&\hspace*{-10mm}
= \left|
1-2i
e^{-i\Delta \tilde E_{31}^{(\mp)} L/2}
|\tilde{U}_{\mu 3}^{(\mp)}|^2
\sin\left(\frac{\Delta \tilde E_{31}^{(\mp)} L}{2}\right)
-2ie^{-i\Delta \tilde E_{21}^{(\mp)} L/2}
|\tilde{U}_{\mu 2}^{(\mp)}|^2
\sin\left(\frac{\Delta \tilde E_{21}^{(\mp)} L}{2}\right)
\right|^2
\nonumber\\
\label{pmm}
\end{eqnarray}
Notice that Eqs.\,(\ref{pme}) and (\ref{pmm}) are exact
and the quantities
$\tilde{U}_{ej}^{(\mp)}\tilde{U}_{\mu j}^{(\mp)\ast}$ and
$|\tilde{U}_{\mu j}^{(\mp)}|^2$ can be exactly
obtained by the formalism by Kimura, Takamura and
Yokomakura (KTY)\,\cite{Kimura:2002hb,Kimura:2002wd}
in the case with constant density of matter,
as long as we know the energy eigenvalues $\tilde{E}_j^{(\mp)}$ exactly.
In reality, however, in order to obtain
$\tilde{E}_j^{(\mp)}$,
we have to use a perturbation method with respect to
$\Delta m^2_{21}/|\Delta m^2_{31}|$.
It should be emphasized that this approximation to
obtain $\tilde{E}_j^{(\mp)}$ is independent of
the baseline length $L$, so even with this approximation,
Eqs.\,(\ref{pme}) and (\ref{pmm}) are valid for arbitrary
baseline length $L$.
As described in Appendix\,\ref{appendixb},
applying the KTY formalism, we obtain
$\tilde{U}_{\alpha j}^{(\mp)}\tilde{U}_{\mu j}^{(\mp)\ast}\,(\alpha=e,\mu; j=2,3)$
to the leading order
in $\Delta m^2_{21}/|\Delta m^2_{31}|$:\,\footnote{
  In the standard parametrization\,\cite{Tanabashi:2018oca} of
  the mixing matrix $U_{\alpha j}$, $U_{\mu 3}$ is real.
  In the KTY formalism, however, the bilinear form
  $\tilde{U}_{\alpha j}^{(\mp)}\tilde{U}_{\beta j}^{(\mp)\ast}$
  in matter is expressed in terms of the same one
  ${U}_{\alpha j}{U}_{\beta j}^{\ast}$ in vacuum, so
  we leave the notation of complex conjugate
  for $U_{\mu 3}$ here to keep generality in
  the parametrization of $U_{\alpha j}$.
}
\begin{eqnarray}
  &{\ }&\hspace{-17mm}
\tilde{U}_{e3}^{(-)}\tilde{U}_{\mu 3}^{(-)\ast}
=U_{e3}U_{\mu 3}^{\ast}
\label{txme3}\\  
&{\ }&\hspace{-17mm}
\tilde{U}_{e3}^{(+)}\tilde{U}_{\mu 3}^{(+)\ast}  
=U_{e3}^{\ast}U_{\mu 3}
\label{txmeb3}\\  
&{\ }&\hspace{-17mm}
\tilde{U}_{e2}^{(-)}\tilde{U}_{\mu 2}^{(-)\ast}
  =\frac{1}{\Delta \tilde{E}_{21}^{(-)}}\left[
\Delta E_{21}U_{e2}U_{\mu 2}^{\ast}
+
\left(\Delta E_{21}
-\Delta \tilde{E}_{21}^{(-)}\right)
\frac{U_{e3}U_{\mu 3}^{\ast}}{2}
\right.
\nonumber\\
&{\ }&\hspace{30mm}
\left.+  A\left(U_{e3}U_{\mu 3}^{\ast}\epsilon_D
+U_{\tau 3}\epsilon_N\right)
\right]
  \label{txme2}\\
&{\ }&\hspace{-17mm}
\tilde{U}_{e2}^{(+)}\tilde{U}_{\mu 2}^{(+)\ast}
  =\frac{1}{\Delta \tilde{E}_{21}^{(+)}}\left[
\Delta E_{21}U_{e2}^{\ast}U_{\mu 2}
+
\left(\Delta E_{21}
-\Delta \tilde{E}_{21}^{(+)}\right)
\frac{U_{e3}^{\ast}U_{\mu 3}}{2}
\right.
\nonumber\\
&{\ }&\hspace{30mm}
\left.-  A\left(U_{e3}^{\ast}U_{\mu 3}\epsilon_D
+U_{\tau 3}\epsilon_N\right)
\right]
  \label{txmeb2}\\
&{\ }&\hspace{-17mm}
|\tilde{U}_{\mu 3}^{(\mp)}|^2=|{U}_{\mu 3}|^2
\label{txmm3}\\  
&{\ }&\hspace{-17mm}
|\tilde{U}_{\mu 2}^{(\mp)}|^2
=\frac{1}{\Delta \tilde{E}_{21}^{(\mp)}}\left\{
 \Delta E_{21} |U_{\mu 2}|^2+
 \left(\Delta E_{21}- \Delta \tilde{E}_{21}^{(\mp)}\right)
 \frac{|U_{\mu 3}|^2}{2}\right\}
\nonumber\\
&{\ }&\hspace{-1mm}
\pm \frac{A}{\Delta \tilde{E}_{21}^{(\mp)}}\left\{
c^2_{13}\left(
1+c_{23}^2-s_{13}^2s_{23}^2
\right)\right.
\nonumber\\
&{\ }&\hspace{17mm}
+2 \, \epsilon_I
+2 \, \epsilon_D
\left(c_{23}^2-s_{13}^2s_{23}^2\right)
\left.+ 2\,\mbox{\rm Re}\left(U_{e3}\epsilon_N\right)
\sin2\theta_{23}
\right\}
\label{txmm2}
\end{eqnarray}
where
$\Delta \tilde{E}_{21}$
is defined by
\begin{eqnarray}
  &{\ }&\hspace*{-15mm}
\Delta \tilde{E}_{21}^{(\mp)}\equiv
\left\{
\left|\Delta E_{21}\cos2\theta_{12}\mp A\left(c_{13}^2 - 2\epsilon_D\right)
\right|^2
    +\left|\Delta E_{21}\sin2\theta_{12}\pm 2A\epsilon_N\right|^2
    \right\}^{1/2}\,,
\label{tdele21}
\end{eqnarray}
and $\epsilon_I$, $\epsilon_D$ and $\epsilon_N$
are defined as
\begin{eqnarray}
  &{\ }&\hspace{-64mm}
  \epsilon_I\equiv\frac{1}{2}\left(
  \epsilon'_{11}+\epsilon'_{22}
  \right)
  \label{epsiloni}\\
  &{\ }&\hspace{-64mm}
\epsilon_D\equiv\frac{1}{2}\left(
  \epsilon'_{22}-\epsilon'_{11}
  \right)=\sum_{f=e,u,d}
\frac{N_f}{N_e}\epsilon_D^{f}
  \label{epsilond}\\
  &{\ }&\hspace{-64mm}
\epsilon_N\equiv\epsilon'_{12}=\sum_{f=e,u,d}
\frac{N_f}{N_e}\epsilon_N^{f}\,.
  \label{epsilonn}
\end{eqnarray}
From Eqs.\,(\ref{txme3}) - (\ref{txmm2})
we see that the appearance probabilities
involve only $\epsilon_D$ and $\epsilon_N$
while the disappearance probabilities
also contain $\epsilon_I$, in addition to
$\epsilon_D$ and $\epsilon_N$.
At low energy long baseline experiments
on the Earth, therefore, all the oscillation
probabilities involves only $\epsilon_I$, $\epsilon_D$ and $\epsilon_N$
and not $\epsilon'_{j3}\,(j=1,2,3)$.
Thus they are advantageous in
determining $\epsilon_D$ and $\epsilon_N$
since there are less NSI parameters
which appear in the oscillation probabilities 
compared with the experiments at higher energy ($E\gtrsim$ 1GeV).

\section{Parameter degeneracy in $\delta$, $\epsilon_I$,
  $\epsilon_D$ and $\epsilon_N$}
In the standard three flavor framework, it has been
known\,\cite{BurguetCastell:2001ez,Minakata:2001qm,Fogli:1996pv,Barger:2001yr}
that, even if we know exactly
the appearance and disappearance probabilities
for neutrinos and antineutrinos for a given
neutrino energy and a given baseline length, there
are in general eight-fold degeneracy
in determination of $\delta$, and this is called
parameter degeneracy in neutrino oscillation.
Here we discuss whether parameter degeneracy
can be resolved at low energy long baseline experiments
in the presence of the NSI.
Our treatment here is based on analytical
expressions of the oscillation probabilities
and the experimental errors are not taken
into account.  However, such discussions
give us an insight into the problem of
parameter degeneracy in the presence of the NSI, like
Refs.\,\cite{BurguetCastell:2001ez,Minakata:2001qm,Fogli:1996pv,Barger:2001yr}
did in the standard case.

Since the oscillation probabilities
(\ref{txme3}) - (\ref{txmm2}) are complicated
functions of the NSI parameters, we make
the following assumptions:

\vglue 3mm
\noindent
(i) All the NSI parameters $\epsilon_I$,
$\epsilon_D$ and $\epsilon_N$ are of order
$s_{13}\simeq 0.15$ or smaller than $s_{13}$, and
if the ratio of the next leading term to the
leading one is of order $s_{13}$, then
the contribution of the next leading term
is negligible.

\vglue 1mm
\noindent
(ii) The following expansion is a good approximation:
$\sin\left(\Delta \tilde E_{21}^{(\mp)} L/2\right)\simeq \Delta \tilde E_{21}^{(\mp)} L/2$.

\vglue 3mm
\noindent
The assumption (i) may be almost justified from
the constraints (\ref{eab}).
On the other hand,
in the energy region of the T2HK and T2HKK
experiments (0.3GeV $\lesssim E \lesssim$ 1GeV),
we have
$\Delta \tilde E_{21}^{(\mp)}L\lesssim 0.54$,
and the error of the approximation
$|(\sin x -x)/x|$ for the range
$0<x<0.54$ is less than 0.05.
So in the present approximation
the assumption (ii) is also justified.
From the assumption (ii),
we can expand the argument of the second term
(solar term) in Eqs.\,(\ref{pme}) for both T2HK
(L=295km) and T2HKK (L=1100km):
\begin{eqnarray}
  &{\ }&\hspace{-12mm}
\tilde{U}_{e2}^{(-)}\tilde{U}_{\mu 2}^{(-)\ast}
\sin\left(\frac{\Delta \tilde E_{21}^{(-)} L}{2}\right)
\nonumber\\
&{\ }&\hspace{-16mm}
  \simeq
\frac{\Delta E_{21} L}{2}U_{e2}U_{\mu 2}^{\ast}
+
\left(\frac{\Delta E_{21}L}{2}
-\frac{\Delta \tilde{E}_{21}^{(\mp)}L}{2}\right)
\frac{U_{e3}U_{\mu 3}^{\ast}}{2}
+  \frac{AL}{2}\left(U_{e3}U_{\mu 3}^{\ast}\epsilon_D
+U_{\tau 3}\epsilon_N\right)
\nonumber\\
\label{solarterm1}\\
&{\ }&\hspace{-16mm}
\tilde{U}_{e2}^{(+)}\tilde{U}_{\mu 2}^{(+)\ast}
\sin\left(\frac{\Delta \tilde E_{21}^{(+)} L}{2}\right)
\nonumber\\
&{\ }&\hspace{-16mm}
  \simeq
\frac{\Delta E_{21} L}{2}U_{e2}^{\ast}U_{\mu 2}
+
\left(\frac{\Delta E_{21}L}{2}
-\frac{\Delta \tilde{E}_{21}^{(\mp)}L}{2}\right)
\frac{U_{e3}^{\ast}U_{\mu 3}}{2}
-  \frac{AL}{2}\left(U_{e3}^{\ast}U_{\mu 3}\epsilon_D
+U_{\tau 3}\epsilon_N\right)
\nonumber\\
\label{solarterm2}\\
  &{\ }&\hspace{-11mm}
\left|\tilde{U}_{\mu 2}^{(\mp)}\right|^2
\sin\left(\frac{\Delta \tilde E_{21}^{(\mp)} L}{2}\right)
\nonumber\\
&{\ }&\hspace{-15mm}
\simeq
 \frac{\Delta E_{21} L}{2} |U_{\mu 2}|^2+
 \left(\frac{\Delta E_{21}L}{2}
 - \frac{\Delta \tilde{E}_{21}^{(\mp)}L}{2}\right)
 \frac{|U_{\mu 3}|^2}{2}
\pm \frac{AL}{2}\left\{
c^2_{13}\left(
1+c_{23}^2-s_{13}^2s_{23}^2
\right)\right.
\nonumber\\
&{\ }&\hspace{-10mm}
+2 \, \epsilon_I
+2 \, \epsilon_D
\left(c_{23}^2-s_{13}^2s_{23}^2\right)
\left.+ 2\,\mbox{\rm Re}\left(U_{e3}\epsilon_N\right)
\sin2\theta_{23}
\right\}
\label{solarterm3}
\end{eqnarray}

First, let us discuss the
disappearance probabilities at the T2HK experiment.
In the case of T2HK ($L$=295km, $E\simeq$ 0.6GeV),
the term $\Delta \tilde E_{21}^{(\mp)} L/2$
on the right hand side of Eq.\,(\ref{pmm})
is of order ($\sim s_{13}^2$), so
the third term on the right hand side of Eq.\,(\ref{pmm})
can be ignored.
Because of the condition (\ref{txmm3})
and because $\Delta \tilde{E}_{31}^{(\mp)}\sim \Delta E_{31}$
to the leading order in $\Delta m^2_{21}/|\Delta m^2_{31}|$,
the disappearance probabilities are reduced
to those in the standard case:
\begin{eqnarray}
  &{\ }&\hspace*{-20mm}
P(\nu_{\mu}\to\nu_{\mu})
=P(\bar{\nu}_{\mu}\to \bar{\nu}_\mu)
=\left|
1-2ie^{-i\Delta  E_{31}L/2}
|{U}_{\mu 3}|^2
\sin\left(\frac{\Delta E_{31} L}{2}\right)
\right|^2
\nonumber\\
&{\ }&\hspace*{29mm}
\simeq 1-\sin^22\theta_{23}\sin^2\left(\frac{\Delta E_{31} L}{2}\right)
\nonumber
\end{eqnarray}
From this, we can determine the value of
$\sin^22\theta_{23}$ in the present approximation.

Next, let us discuss the
appearance probabilities of T2HK.
Since the second and third terms
on the right hand side of Eq.\,(\ref{solarterm1}) are multiplied by
small quantities such as $U_{e3}=e^{-i\delta}s_{13}$
and $\epsilon_N$,
the only surviving term on the right hand side of
Eq.\,(\ref{solarterm1}) is the first one
$U_{e2}U_{\mu 2}^{\ast}\Delta E_{21} L/2$.
Thus, in the present approximation
in which terms higher than $s_{13}$ etc.
are ignored, the problem
of determination of $\delta$ at T2HK
is reduced to the same problem as that in the
standard three flavor framework.
Since the baseline length of T2HK
satisfies $|\Delta E_{31}| L/2\simeq \pi/2$
and the mass hierarchy has a ratio
$\Delta m^2_{21}/|\Delta m^2_{31}|\simeq 1/30$, we have
\begin{eqnarray}
  &{\ }&\hspace{-26mm}
P(\nu_{\mu}\to\nu_{e})
\simeq 4 \left|
U_{e3}U_{\mu 3}^{\ast}
\sin\left(\frac{\Delta E_{31} L}{2}\right)
+e^{i\Delta E_{31} L/2}
\frac{\Delta E_{21} L}{2}U_{e2}U_{\mu 2}^{\ast}
\right|^2
\nonumber\\
&{\ }&\hspace*{-4mm}
\simeq \left|\mbox{\rm sign}\left(\Delta m^2_{31}\right)
\left(2e^{-i\delta}s_{13}s_{23}
+i\frac{\pi}{4}\cdot
\frac{\Delta m^2_{21}}{|\Delta m^2_{31}|}\cdot
c_{23}\sin2\theta_{12}\right)
\right|^2
\nonumber\\
&{\ }&\hspace*{-4mm}
\simeq \left|2e^{-i\delta}s_{13}s_{23}
+i\frac{\pi}{120}
c_{23}\sin2\theta_{12}
\right|^2
\label{pme-t2hk}
\end{eqnarray}
\begin{eqnarray}
&{\ }&\hspace*{-26mm}
  P(\bar{\nu}_{\mu}\to \bar{\nu}_e)
\simeq 4 \left|
U_{e3}U_{\mu 3}^{\ast} 
\sin\left(\frac{\Delta E_{31} L}{2}\right)
+e^{-i\Delta E_{31} L/2}
\frac{\Delta E_{21} L}{2}U_{e2}U_{\mu 2}^{\ast}
\right|^2
\nonumber\\
&{\ }&\hspace*{-4mm}
\simeq \left|\mbox{\rm sign}\left(\Delta m^2_{31}\right)
\left(2e^{-i\delta}s_{13}s_{23}
-i\frac{\pi}{4}\cdot
\frac{\Delta m^2_{21}}{|\Delta m^2_{31}|}\cdot
c_{23}\sin2\theta_{12}\right)
\right|^2
\nonumber\\
&{\ }&\hspace*{-4mm}
\simeq  \left|
2e^{-i\delta}s_{13}s_{23}
-i\frac{\pi}{120}
c_{23}\sin2\theta_{12}
\right|^2
\label{pmeb-t2hk}
\end{eqnarray}
Notice that the appearance probabilities
(\ref{pme-t2hk}) and (\ref{pmeb-t2hk})
at T2HK are independent not only of
the NSI parameters but also
of the mass hierarchy
(sign($\Delta m^2_{31}$))
in the present approximation.
This implies that there is no way to determine the
mass hierarchy from the T2HK appearance channel,
as is well known.
The T2HK experiment as well as T2K\,\cite{Itow:2001ee}
is performed
at the oscillation maximum ($|\Delta E_{31}| L/2\simeq \pi/2$),
and it is known\,\cite{Barger:2001yr} that
the so-called intrinsic degeneracy becomes
the ambiguity in the sign of $\cos\delta$ in this case.
This ambiguity cannot be removed by the T2HK alone,
and as we will see below, we need the T2HKK data to
remove this ambiguity.
On the other hand, the appearance probabilities have some dependence
on the octant of $\theta_{23}$, and we can resolve
the octant degeneracy.

\begin{figure}[htbp]
 \begin{minipage}{0.32\hsize}
\hspace*{3mm}
  \includegraphics[width=3.5in]{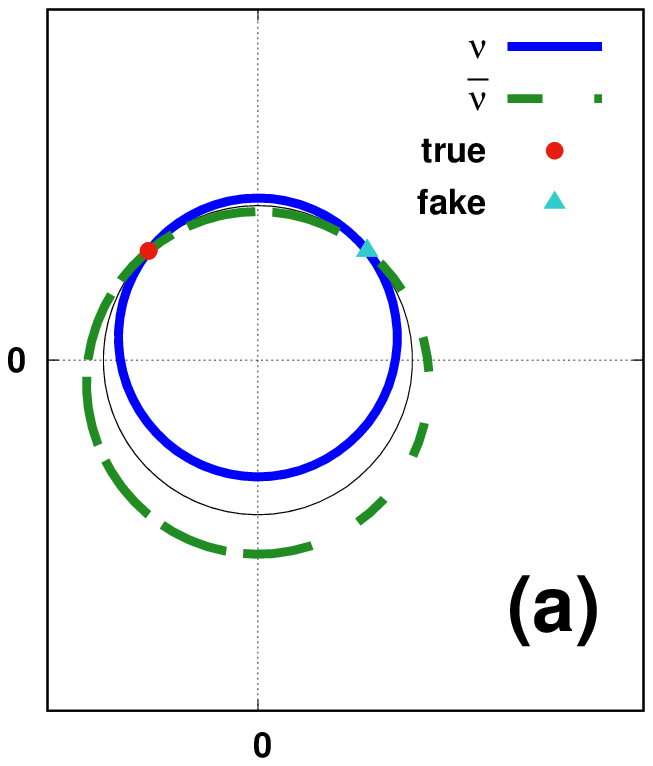}
 \end{minipage}
 \begin{minipage}{0.32\hsize}
\hspace*{-8mm}
   \includegraphics[width=3.5in]{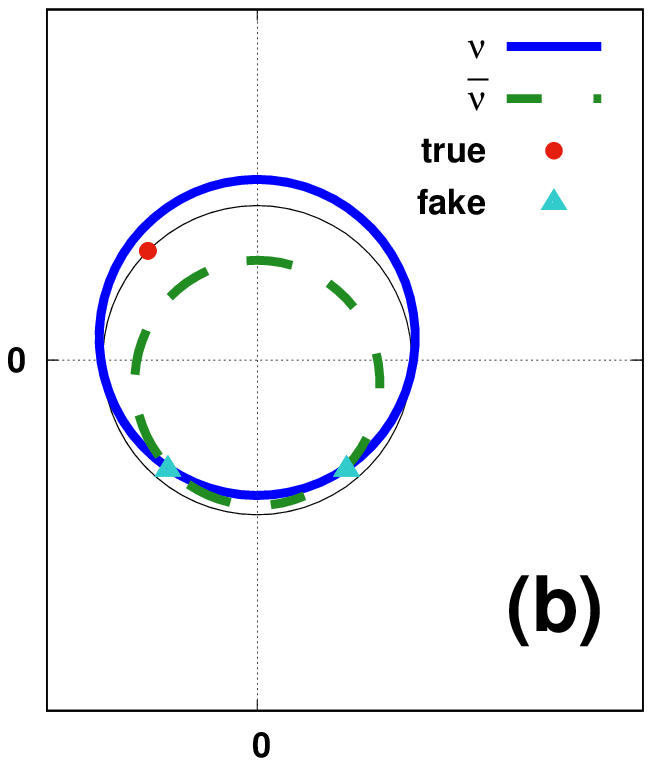}
 \end{minipage}
 \begin{minipage}{0.32\hsize}
   \includegraphics[width=3.5in]{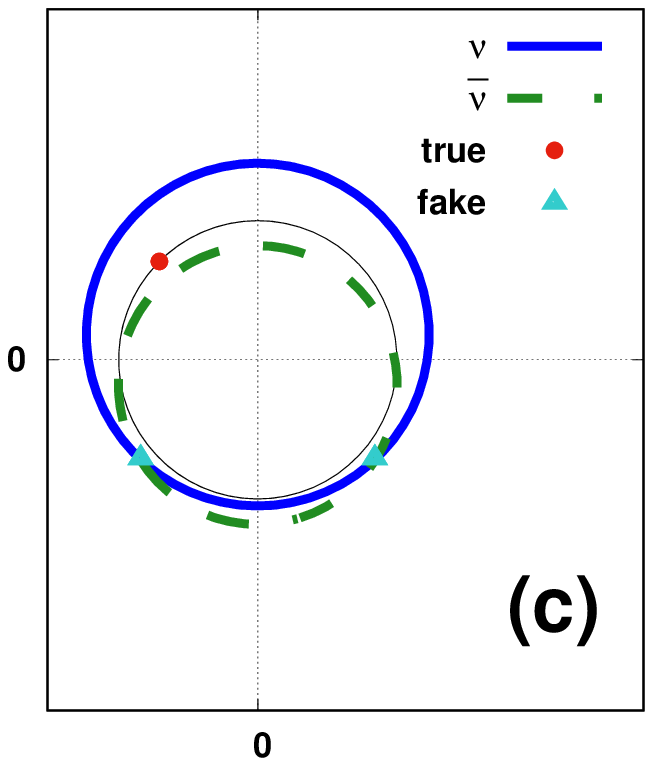}
 \end{minipage}
  \vspace*{9mm}
  \caption{Determination of $\delta$ at T2HK
    using the complex plane of $z\equiv 2e^{-i\delta}s_{13}s_{23}$
    in the case where the true value is $\delta^{\mbox{\tiny\rm true}}=5\pi/4$.
    The thick solid (dashed) circle stands for
    $P(\nu_{\mu}\to\nu_{e};\delta,\theta_{23})=
    P(\nu_{\mu}\to\nu_{e};\delta^{\mbox{\tiny\rm true}},\theta_{23}^{\mbox{\tiny\rm true}})$
($P(\bar{\nu}_{\mu}\to \bar{\nu}_e;\delta,\theta_{23})=
    P(\bar{\nu}_{\mu}\to \bar{\nu}_e;\delta^{\mbox{\tiny\rm true}},
    \theta_{23}^{\mbox{\tiny\rm true}}$),
    while the thin solid circle stands for
    the circle with a radius $2s_{13}s_{23}$.
    (a): The case with the right octant
    ($\theta_{23}^{\mbox{\tiny\rm true}}=16\pi/60$,
    $\delta^{\mbox{\tiny\rm true}}=5\pi/4$).
    The true (fake) point $2\exp(-5i\pi/4)s_{13}s_{23}$
    ($2\exp(-7i\pi/4)s_{13}s_{23}$) with $\delta=5\pi/4$
    ($\delta=7\pi/4$) is depicted
    by a filled circle (a filled triangle).
    From the appearance channel, only
    $\sin\delta$ is determined, leaving
    the sign of $\cos\delta$ unknown.
    (b): The wrong octant $\theta_{23}=14\pi/60<\pi/4$
    in the case where the true value
    $\theta_{23}^{\mbox{\tiny\rm true}}=16\pi/60$
    is in the higher octant.
    (c): The wrong octant $\theta_{23}=16\pi/60>\pi/4$
    in the case where the true value 
    $\theta_{23}^{\mbox{\tiny\rm true}}=14\pi/60$
    is in the lower octant.
    With the wrong octant, the solution for Eqs.\,(\ref{pme-t2hk2}) and
    (\ref{pmeb-t2hk2}) is inconsistent with the condition
    $|2e^{-i\delta}s_{13}s_{23}|=2s_{13}s_{23}$, i.e., the intersection of
    the two thick circles is not on the thin circle.
    }
  \label{circle-t2hk}
\end{figure}

Here, for concreteness, we take the
true values as $\delta^{\mbox{\tiny\rm true}}=5\pi/4$ and
$\theta_{23}^{\mbox{\tiny\rm true}}=16\pi/60$ (with
$(s_{23}^{\mbox{\tiny\rm true}})^2=0.552$)
which are almost the best fit values
at present~\cite{Tanabashi:2018oca},
respectively.
The problem of determining $\delta$ from
the two equations
\begin{eqnarray}
  &{\ }&\hspace{-10mm}
  P\left(\nu_{\mu}\to\nu_{e};\delta,
  \theta_{23}=\frac{\pi}{4}\pm\frac{\pi}{60}\right)=
P\left(\nu_{\mu}\to\nu_{e};\delta^{\mbox{\tiny\rm true}}=\frac{5}{4}\pi,
\theta_{23}^{\mbox{\tiny\rm true}}=\frac{16}{60}\pi\right)
\label{pme-t2hk2}
\end{eqnarray}
and
\begin{eqnarray}
  &{\ }&\hspace{-10mm}
  P\left(\bar{\nu}_{\mu}\to \bar{\nu}_e;\delta,
  \theta_{23}=\frac{\pi}{4}\pm\frac{\pi}{60}\right)=
P\left(\bar{\nu}_{\mu}\to \bar{\nu}_e
;\delta^{\mbox{\tiny\rm true}}=\frac{5}{4}\pi,
\theta_{23}^{\mbox{\tiny\rm true}}=\frac{16}{60}\pi\right)
\label{pmeb-t2hk2}
\end{eqnarray}
can be solved by looking for the
intersection between the two circles
in the complex plane of the variable
$z\equiv 2\exp(-i\delta)s_{13}s_{23}$
as in Fig.\ref{circle-t2hk}.
Eq.\,(\ref{pme-t2hk2}) ((\ref{pmeb-t2hk2}))
tells us that
the distance between the points
$2\exp(-i\delta)s_{13}s_{23}$
and $-i(\pi/120)c_{23}\sin2\theta_{12}$
($i(\pi/120)c_{23}\sin2\theta_{12}$)
in the complex plane
is the same as that between the points
$2\exp(-5i\pi/4)s_{13}s_{23}$
and $i(\pi/120)c_{23}\sin2\theta_{12}$
($i(\pi/120)c_{23}\sin2\theta_{12}$),
respectively.
If our hypothesis on the octant of $\theta_{23}$
is correct
(in the present case it is in the higher octant
($\theta_{23}^{\mbox{\tiny\rm true}}=16\pi/60>\pi/4$)), then
we have two solutions corresponding to
$\cos\delta=\pm|\cos\delta|$, as is shown in
Fig.\,\ref{circle-t2hk}\,(a).
On the other hand, if our hypothesis on the octant of $\theta_{23}$
is wrong, then the absolute value of the intersection
points is not equal to $2s_{13}s_{23}$
(Fig.\,\ref{circle-t2hk}\,(b) where
a fit with $\theta_{23}=14\pi/60<\pi/4$ is attempted
for the true value $\theta_{23}^{\mbox{\tiny\rm true}}=16\pi/60$)
or 
(Fig.\,\ref{circle-t2hk}\,(c) where
a fit with $\theta_{23}=16\pi/60>\pi/4$ is attempted
for the true value $\theta_{23}^{\mbox{\tiny\rm true}}=14\pi/60$),
and we can reject the wrong hypotheses
on the assumption
that difference between the true and fake
points is large enough compared with the
experimental errors.
Note that the precise value of $\theta_{13}$,
which was determined by the reactor
experiments\,\cite{Tanabashi:2018oca}, is crucial
to resolve the octant degeneracy
because it uniquely specifies the
radius of the thin circle
in Fig.\,\ref{circle-t2hk}.

To summarize so far, we have the following
results from the T2HK data:

\noindent
$\bullet$ For the sign degeneracy and
the NSI parameters, we do not
get any information.

\noindent
$\bullet$ For the intrinsic degeneracy,
we can determine the value of $\sin\delta$
but we still have ambiguity in the sign
of $\cos\delta$.

\noindent
$\bullet$ For the octant degeneracy,
we can resolve it, on the assumption
that deviation $|\pi/4-\theta_{23}|$
is large enough compared with the
experimental errors.

Let us now turn to the appearance probabilities
at T2HKK ($L$=1100km, 0.3GeV$\lesssim E\lesssim$ 1.1GeV).
Since the T2HK appearance channel
enables us to determine the value of
$\sin\delta$ and the octant of $\theta_{23}$,
we assume in the following discussions that
we know the value of $\sin\delta$ and $\theta_{23}$,
and the unknown are sign($\cos\delta$),
sign($\Delta m_{31}^2$) and the NSI parameters.
In the case of T2HKK, while $AL/2\,(\simeq 1/4)$ and
$\Delta E_{21} L\,(\sim 0.2\,(0.6\mbox{\rm GeV}/E))$
can no longer be treated as small quantity,
the term $U_{e3}U_{\mu 3}^{\ast}\epsilon_D$
in Eq.\,(\ref{solarterm1})
is of order $s_{13}^2$ from our
assumption, so it can be ignored.
Eq.\,(\ref{solarterm1}) contains
the factor $\Delta \tilde{E}_{21}^{(\mp)}$
which is defined in Eq.\,(\ref{tdele21}),
and it has the following expansion
with respect to the small NSI parameters:
\begin{eqnarray}
  &{\ }&\hspace*{-15mm}
\Delta \tilde{E}_{21}^{(\mp)}\simeq
\left. \Delta \tilde{E}_{21}^{(\mp)}\right|_{\mbox{\footnotesize\rm std}}
+\delta\Delta \tilde{E}_{21}^{(\mp)}
\label{tdele21v2}\\
  &{\ }&\hspace*{-15mm}
  \left.\Delta \tilde{E}_{21}^{(\mp)}\right|_{\mbox{\footnotesize\rm std}}
  \equiv
\left\{
\left(\Delta E_{21}\cos2\theta_{12}\mp A c_{13}^2\right)^2
    +\left(\Delta E_{21}\sin2\theta_{12}\right)^2
    \right\}^{1/2}\,
 \label{tdele21v3}\\
&{\ }&\hspace*{-15mm}
\delta\Delta \tilde{E}_{21}^{(\mp)}
\equiv\pm \frac{2A}{\left.\Delta \tilde{E}_{21}^{(\mp)}\right|_{\mbox{\footnotesize\rm std}}}
\left\{\epsilon_D\left(\Delta E_{21}\cos2\theta_{12}\mp Ac_{13}^2 \right)
  + \mbox{\rm Re}(\epsilon_N) \Delta E_{21}\sin2\theta_{12}\right\}\,.
  \label{tdele21v4}
\end{eqnarray}
This small correction $\delta\Delta \tilde{E}_{21}^{(\mp)}$ also
gives a contribution to the appearance probabilities,
and we have
\begin{eqnarray}
  &{\ }&\hspace{-1mm}
  P(\nu_{\mu}\to\nu_{e})
\nonumber\\
\nonumber\\
&{\ }&\hspace*{-4mm}
\simeq \left|
2U_{e3}U_{\mu 3}^{\ast}
\sin\left(\frac{\Delta E_{31} L}{2}\right)
+e^{i\Delta E_{31} L/2}
\left\{
\Delta E_{21} L\, U_{e2}U_{\mu 2}^{\ast}
\right.\right.
\nonumber\\
&{\ }&\hspace*{-0mm}
\left.\left.+\left(\Delta E_{21}L
-\left.\Delta \tilde{E}_{21}^{(-)}L\right|_{\epsilon_{\alpha\beta}=0}\right)
\frac{U_{e3}U_{\mu 3}^{\ast}}{2}
+  AL\,U_{\tau 3}\epsilon_N
\right\}
\right|^2
\nonumber\\
&{\ }&\hspace*{-4mm}
\simeq \left|
\Delta E_{21} L\, U_{e2}U_{\mu 2}^{\ast}
+\left\{
2e^{-i\Delta E_{31} L/2}
\sin\left(\frac{\Delta E_{31} L}{2}\right)
+\frac{\Delta E_{21}L}{2}
-\left.\frac{\Delta \tilde{E}_{21}^{(-)}L}{2}\right|_{\epsilon_{\alpha\beta}=0}
\right\}U_{e3}U_{\mu 3}^{\ast}\right.
\nonumber\\
&{\ }&\hspace*{4mm}
\left.+  AL\,U_{\tau 3}\epsilon_N
\right|^2
\label{pme-t2hkk}\\
\nonumber\\
&{\ }&\hspace*{-1mm}
  P(\bar{\nu}_{\mu}\to \bar{\nu}_e)
\nonumber\\
&{\ }&\hspace*{-4mm}
\simeq \left|
\Delta E_{21} L\, U_{e2}^{\ast}U_{\mu 2}
+\left\{
2e^{-i\Delta E_{31} L/2}
\sin\left(\frac{\Delta E_{31} L}{2}\right)
+\frac{\Delta E_{21}L}{2}
-\left.\frac{\Delta \tilde{E}_{21}^{(+)}L}{2}\right|_{\epsilon_{\alpha\beta}=0}
\right\}U_{e3}^{\ast}U_{\mu 3}\right.
\nonumber\\
&{\ }&\hspace*{4mm}
\left.-  AL\,U_{\tau 3}\epsilon_N
\right|^2
\label{pmeb-t2hkk}
\end{eqnarray}
In the last equation in Eq.\,(\ref{pme-t2hkk}),
the first line, which is assumed to be known
up to the sign of $\cos\delta$,
is the contribution of the standard three flavor framework
and the second line is the
NSI contribution.
Assuming that $\delta$ is already known
from the T2HK data (up to the sign of $\cos\delta$),
the two equations
\begin{eqnarray}
  &{\ }&\hspace{-70mm}
  P(\nu_{\mu}\to\nu_{e};\epsilon_N)=
  P(\nu_{\mu}\to\nu_{e};\epsilon_N^{\mbox{\tiny\rm true}})
\label{pme-t2hkk2}
\end{eqnarray}
and
\begin{eqnarray}
  &{\ }&\hspace{-70mm}
P(\bar{\nu}_{\mu}\to \bar{\nu}_e;\epsilon_N)=
P(\bar{\nu}_{\mu}\to \bar{\nu}_e;\epsilon_N^{\mbox{\tiny\rm true}})
\label{pmeb-t2hkk2}
\end{eqnarray}
give us a condition on $\epsilon_N$.

A remark is in order.
As was emphasized in Ref.\,\cite{Ge:2016dlx},
the reason that information on
$\epsilon_N$ can be still obtained
after expanding a sine function
with a small argument as
$\sin(\Delta \tilde E_{21}^{(\mp)} L/2)\simeq
\Delta \tilde E_{21}^{(\mp)} L/2$
is because this is the case where
a so-called
vacuum mimicking phenomenon
\cite{Wolfenstein:1977ue,DeRujula:1998umv,Freund:1999gy,Akhmedov:2000cs,Lipari:2001ds,Minakata:2000ee,Minakata:2000wm,Yasuda:2001va}
does not occur.
In the standard three flavor framework,
if the argument of a sine function
is small and expanded as $\sin x\simeq x$,
then the oscillation probability in matter
is reduced to the one in vacuum, and
this is call a vacuum mimicking phenomenon.
In the present case, however,
even after the approximation
$\sin(\Delta \tilde E_{21}^{(\mp)} L/2)\simeq
\Delta \tilde E_{21}^{(\mp)} L/2$
is used, the term with $\epsilon_N$
remains.  This is an advantage of
a long baseline experiment ($L\gtrsim$1000km)
at low energy ($E\lesssim$1GeV),
such as T2HKK, since the other NSI parameters
do not appear in the appearance probability
to the leading order at low energy.

\begin{figure}[!h]
  \includegraphics[width=3.3in]{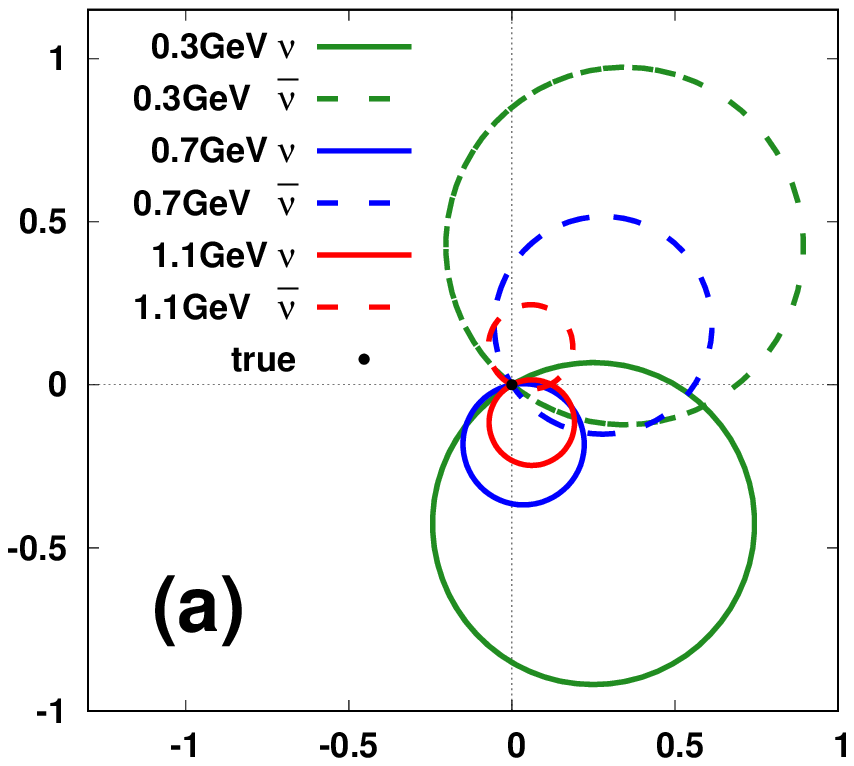}
  \includegraphics[width=3.3in]{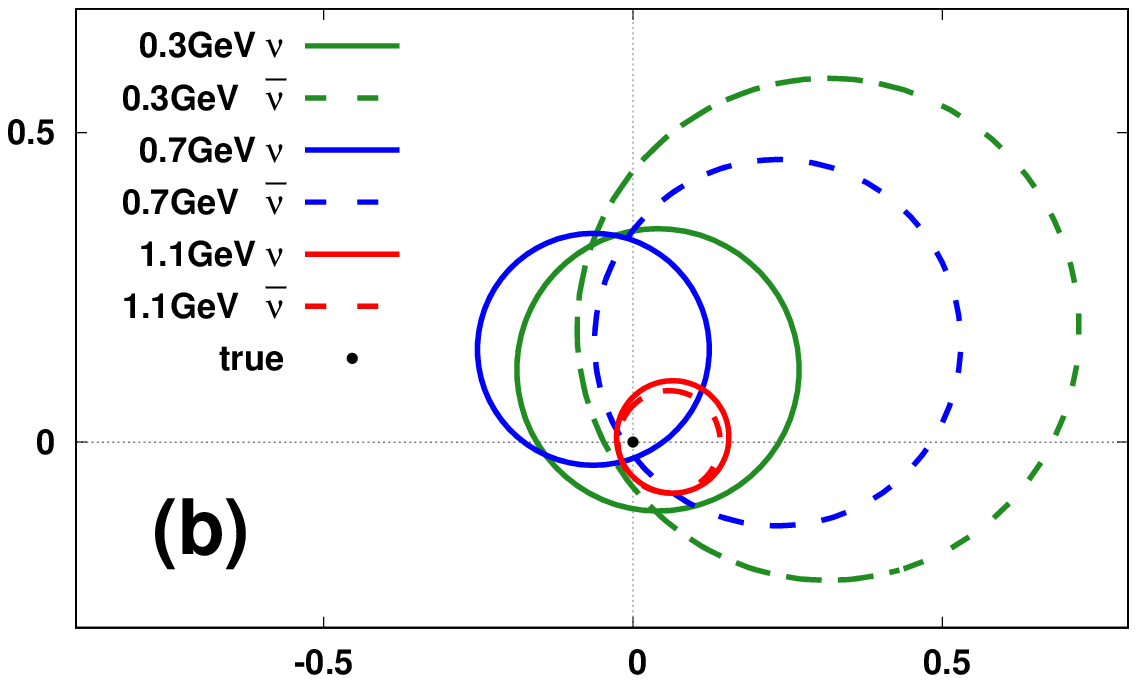}
  \includegraphics[width=3.3in]{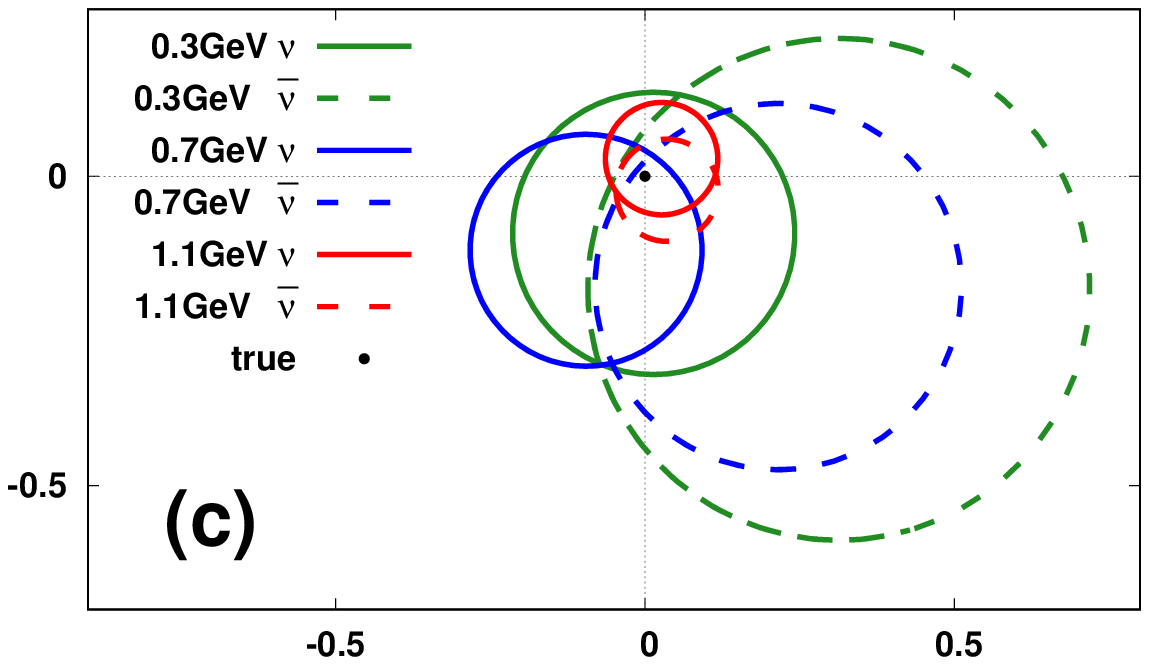}
  \includegraphics[width=3.3in]{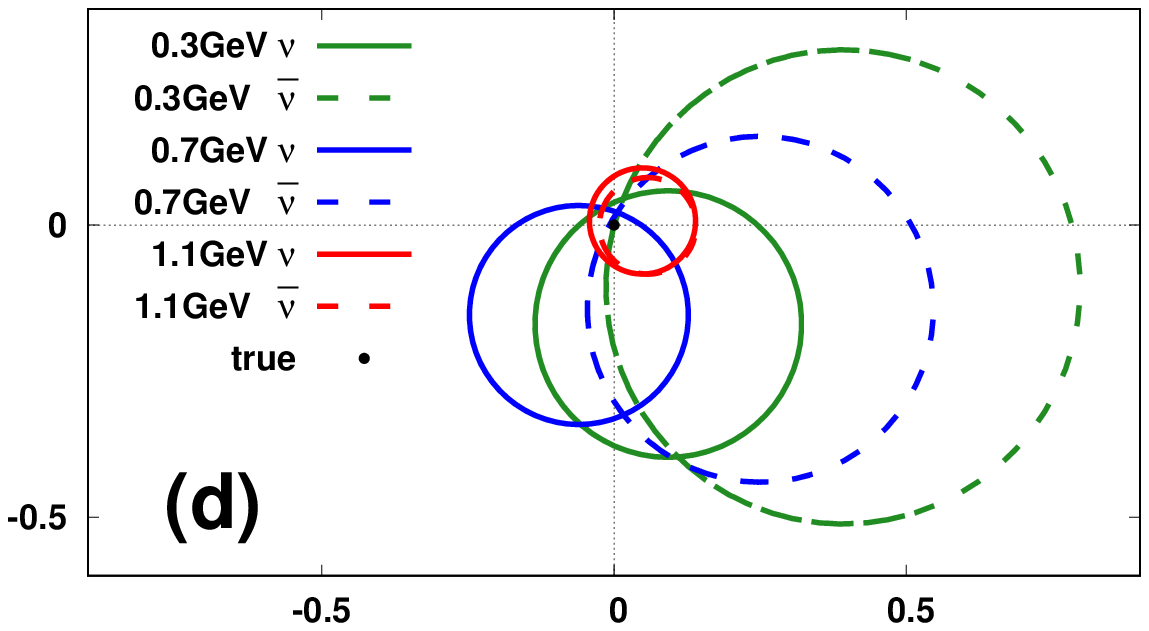}
  \vspace*{5mm}
  \caption{Four possible patterns at T2HKK in the
    complex plane of $z\equiv (AL/2)U_{\tau 3}\epsilon_N
    \simeq 0.18\,\epsilon_N$.  The true solution can be
    selected by demanding that the three pairs of the
    circles has a common intersection.
    (a): a solution with choice of
    right sign($\sin\delta$) and right sign($\Delta m_{31}^2$).
    (b): a solution with choice of
    wrong sign($\sin\delta$) and right sign($\Delta m_{31}^2$).
    (c): a solution with choice of
    right sign($\sin\delta$) and wrong sign($\Delta m_{31}^2$).
    (d): a solution with choice of
    wrong sign($\sin\delta$) and wrong sign($\Delta m_{31}^2$).}
\label{circle-t2hkk}
\end{figure}

As in the case of T2HK, Eqs.\,(\ref{pme-t2hkk2})
and (\ref{pmeb-t2hkk2})
represent two circles in the complex plane
of $z\equiv AL\,U_{\tau 3}\epsilon_N$,
and in general there are two intersections.
To reject the fake solutions, we need more information.
We therefore consider the appearance probabilities
at different three energy regions, e.g.,
$E$=0.3 GeV, $E$=0.7 GeV and $E$=1.1 GeV.
Here we take $\epsilon_N^{\mbox{\tiny\rm true}}=0$ as
the true value for simplicity.
As we see in Fig.\,\ref{circle-t2hkk},
there are four possible cases
with right/wrong sign of $\cos\delta$
and right/wrong sign of $\Delta m_{31}^2$.
By demanding that there be a common
intersection point among the three
pairs of circles, we can
resolve degeneracy of sign($\sin\delta$)
and that of sign($\Delta m_{31}^2$), and
we can determine both Re($\epsilon_N$) and
Im($\epsilon_N$), on the assumption
that the difference between the true
and fake points is large enough compared with the
experimental errors.
So far we have taken $\epsilon_N^{\mbox{\tiny\rm true}}=0$ as
the true value for simplicity.
If the true value $\epsilon_N^{\mbox{\tiny\rm true}}$ is nonzero,
then the same argument can
be applied, since all the positions of
the circles and $\epsilon_N^{\mbox{\tiny\rm true}}$
in the complex plane are shifted
by $\epsilon_N^{\mbox{\tiny\rm true}}\,(\ne 0$).
Hence even for $\epsilon_N^{\mbox{\tiny\rm true}} \ne 0$,
we can determine $\epsilon_N$
from the appearance probabilities
of T2HKK and all the information from T2HK.

Finally, let us discuss determination of $\epsilon_D$.
In our approximation,
$\epsilon_D$ does not appear in the appearance
probabilities to the leading order.
So far we have already determined $\delta$,
$\theta_{23}$ and $\epsilon_N$, so we assume
in the following discussions that we already
know the value of these parameters.
To get information on $\epsilon_D$,
let us discuss the disappearance probabilities at T2HKK.
They are given by (See Appendix\,\ref{appendixc} for details.)
\begin{eqnarray}
&{\ }&\hspace{-34mm}
  \left\{\begin{array}{l}
  P(\nu_{\mu}\to\nu_{\mu})\\
P(\bar{\nu}_{\mu}\to \bar{\nu}_\mu)  
  \end{array}\right\}
\simeq  \left|
AL\,\epsilon_I+f^{(\mp)}\epsilon_D
+g^{(\mp)}
\right|^2\,,
\label{pmm1}
\end{eqnarray}
where $f^{(\mp)}=O(1)$ and $g^{(\mp)}=O(1)$ are defined by
\begin{eqnarray}
 &{\ }&\hspace{-20mm}
 f^{(\mp)}\equiv
 AL\left(
 c_{23}^2-i\frac{\Delta E_{21}\cos2\theta_{12}\mp Ac_{13}^2}
 {\left.\Delta \tilde{E}_{21}^{(\mp)}\right|_{\mbox{\footnotesize\rm std}}}{\cal F}
 \right)
\label{fmp}\\
 &{\ }&\hspace{-20mm}
 g^{(\mp)}\equiv
 \pm\exp\left(i\frac{L}{2}\left.
 \Delta \tilde E_{21}^{(\mp)}\right|_{\mbox{\tiny\rm std}} \right)
\left\{\frac{i}{2}
+e^{-i\Delta E_{31} L/2}
|{U}_{\mu 3}|^2
\sin\left(\frac{\Delta E_{31} L}{2}\right)\right\}
\nonumber\\
&{\ }&\hspace*{-8mm}
\pm   \frac{\Delta E_{21} L}{2} |U_{\mu 2}|^2
\pm \left(\frac{L}{2}\Delta E_{21}
 - \frac{L}{2}\left.\Delta \tilde E_{21}^{(\mp)}\right|_{\mbox{\tiny\rm std}} \right)
 \frac{|U_{\mu 3}|^2}{2}+ \frac{AL}{2}\left(
 1+c_{23}^2\right)
 \nonumber\\
&{\ }&\hspace*{-8mm}
 -iAL\mbox{\rm Re}(\epsilon_N)
 \frac{\Delta E_{21}\sin2\theta_{12}}
 {\left.\Delta \tilde{E}_{21}^{(\mp)}\right|_{\mbox{\footnotesize\rm std}}}{\cal F}
\label{gmp}
\end{eqnarray}
with
\begin{eqnarray}
&{\ }&\hspace{-20mm}
{\cal F}\equiv   \frac{\Delta E_{21} L}{2} |U_{\mu 2}|^2+
 \left(\frac{L}{2}\Delta E_{21}
 - \frac{L}{2}\left.\Delta \tilde E_{21}^{(\mp)}\right|_{\mbox{\tiny\rm std}}
 -i\right)
 \frac{|U_{\mu 3}|^2}{2}\,.
 \label{f}
\end{eqnarray}

From the discussions on the
T2HK data and the appearance channel of T2HKK,
we already know the values of
$\delta$ and $\epsilon_N$.
Assuming that the true values of the NSI parameters
$\epsilon_N^{\mbox{\tiny\rm true}}$, $\epsilon_I^{\mbox{\tiny\rm true}}$
and $\epsilon_D^{\mbox{\tiny\rm true}}$
are zero for simplicity, the following equations
give us information on $\epsilon_I$
and $\epsilon_D$:
\begin{eqnarray}
  &{\ }&\hspace{-30mm}
P(\nu_{\mu}\to\nu_{\mu};\epsilon_I, \epsilon_D)=P(\nu_{\mu}\to\nu_{\mu};0,0)
\nonumber\\
&{\ }&\hspace{-40mm}
\left|
AL\,\epsilon_I+f^{(-)}\epsilon_D
+g^{(-)}
\right|^2=\left| g^{(-)}\right|^2
\label{pmm2}\\
\nonumber\\
  &{\ }&\hspace{-30mm}
P(\bar{\nu}_{\mu}\to \bar{\nu}_\mu;\epsilon_I, \epsilon_D)=
P(\bar{\nu}_{\mu}\to \bar{\nu}_\mu;0,0)
\nonumber\\
&{\ }&\hspace{-40mm}
\left|
AL\,\epsilon_I+f^{(+)}\epsilon_D
+g^{(+)}
\right|^2=\left| g^{(+)}\right|^2
\label{pmmb2}
\end{eqnarray}
Unlike in the case of the
appearance probabilities,
where the contributions from
the atmospheric oscillation
and from the solar one are both small,
the NSI contributions in Eq.\,(\ref{pmmv3})
are small compared with the one
from atmospheric oscillation.
So we can expand the disappearance
probabilities in term of the small
parameters $\epsilon_I$ and $\epsilon_D$.
\begin{eqnarray}
&{\ }&\hspace{-34mm}
  AL\,\mbox{\rm Re}\left[g^{(-)}\right]\,\epsilon_I
  +\mbox{\rm Re}\left[g^{(-)}f^{(-)\ast}\right]\,\epsilon_D
=0
\label{pmm3}
\end{eqnarray}
\begin{eqnarray}
&{\ }&\hspace{-34mm}
  AL\,\mbox{\rm Re}\left[g^{(+)}\right]\,\epsilon_I
  +\mbox{\rm Re}\left[g^{(+)}f^{(+)\ast}\right]\,\epsilon_D
=0
\label{pmmb3}
\end{eqnarray}
From these two equations we can determine
$\epsilon_I$ and $\epsilon_D$.
Here we have assumed that the true value of
the NSI parameters are zero for simplicity,
but even for a nonvanishing value of
the NSI parameters, the same argument
can be applied.

To summarize, we have seen that, 
because the T2HK experiment has
a relatively short baseline length,
the oscillation probabilities at T2HK
are approximately independent of
the NSI parameters and T2HK can
determine the value of $\sin\delta$
and it can resolve the octant degeneracy,
on the assumption
that the difference between the true
and fake points is large enough compared with the
experimental errors.
Furthermore, the T2HKK experiment
can resolve degeneracy of the sign
of $\Delta m_{31}^2$
as well as the ambiguity of
$\cos\delta$.  By combining
the appearance and disappearance
probabilities at T2HK and T2HKK,
we can determine the NSI
parameters $\epsilon_N$,
$\epsilon_I$ and $\epsilon_D$.

\section{Conclusions}
At low energy ($E\lesssim$ 1GeV), the description
in the solar flavor basis is useful.
In particular, in the presence of the
nonstandard interactions in propagation of
neutrinos, assuming that the NSI parameters
are at most of order O($s_{13}$),
the appearance probabilities at low energy
depend approximately only on one ($\epsilon_N$) of the
NSI parameters, while the disappearance
ones do on three (Re($\epsilon_N$),
$\epsilon_I$ and $\epsilon_D$).
Furthermore, assuming that
the experimental errors are small enough
to justify the analytical discussions
on the oscillation probabilities,
we discussed how parameter degeneracy can be resolved
by combining the T2HK and T2HKK experiments.
These two low energy long baseline experiments
are complementary to each other, because
T2HK has little sensitivity to the matter effect
and can therefore determine $\sin\delta$
and the octant of $\theta_{23}$ without
being disturbed by the existence of the NSI
whereas T2HKK has sensitivity to the matter effect
and can give us information on the NSI parameters
as well as sign($\sin\delta$) and sign($\Delta m_{31}^2$).
Our treatment in this work is
qualitative in the sense that the experimental
errors are not taken into account,
and quantitative estimation of the
experimental errors is beyond the scope
of this work.  Nevertheless, we hope
that the present work sheds light on the
advantage of low energy long baseline experiments
to investigate the NSI which is suggested
by the tension between the solar neutrino data and
that from the KamLAND experiment.

\section*{Acknowledgments}
This research was partly supported by a Grant-in-Aid for Scientific
Research of the Ministry of Education, Science and Culture, under
Grants No. 18K03653 and No. 18H05543.

\vglue 10mm
\noindent
{\Large\bf Appendix}

\appendix

\section{Analytical form of the oscillation probability
and the Kimura-Takamura-Yokomakura formalism}
\label{appendixa}
If neutrino has a potential, which can
in general has off diagonal components
in the presence of the NSI, then the
Hamiltonian in matter with constant
density for neutrinos and antineutrinos
can be formally diagonalized as
\begin{eqnarray}
  &{\ }&\hspace*{-30mm}
  \left\{\begin{array}{l}
  U{\cal E}U^{-1}+{\cal A}\\
  U^\ast{\cal E}(U^\ast)^{-1}-{\cal A}
  \end{array}\right\}
=\tilde{U}^{(\mp)}\tilde{{\cal E}}^{(\mp)}(\tilde{U}^{(\mp)})^{-1}\,.
\label{sch1}
\end{eqnarray}
In Eq.\,(\ref{sch1}),
${\cal A}$ is the $3\times 3$ matrix of the
matter potential defined in Eq.\,(\ref{matter-np-solar}),
\begin{eqnarray}
&{\ }&\hspace*{-50mm}
  {\cal E}\equiv{\mbox{\rm diag}}\left(0,\Delta E_{21},\Delta E_{31}\right)
  \label{sch2}
\end{eqnarray}
with
\begin{eqnarray}
&{\ }&\hspace*{-28mm}
  \Delta E_{jk}\equiv E_j - E_k \simeq
\frac{m_j^2-m_k^2}{2E}
  \equiv\frac{\Delta m_{jk}^2}{2E}
\end{eqnarray}
is the diagonal matrix with the energy eigenvalue of
each mass eigenstate where the
identity matrix times $E_1$ was subtracted without
affecting the oscillation probability, and
\begin{eqnarray}
&{\ }&\hspace*{-51mm}
\tilde{\cal E}^{(\mp)}\equiv{\mbox{\rm diag}}\left(
\tilde{E}_1^{(\mp)},\tilde{E}_2^{(\mp)},\tilde{E}_3^{(\mp)}\right)
\label{sch3}
\end{eqnarray}
is the diagonal matrix with the energy eigenvalue 
in matter.
From Eq.\,(\ref{sch1}) one can obtain the
oscillation probability
\begin{eqnarray}
&{\ }&\hspace*{-6mm}
  \left\{\begin{array}{l}
  P(\nu_\alpha\rightarrow\nu_\beta)\\
  P(\bar{\nu}_\alpha\rightarrow\bar{\nu}_\beta)
\end{array}\right\}
\nonumber\\
&{\ }&\hspace*{-10mm}
=\left|\sum_{j=1}^3
\tilde{U}_{\beta j}^{(\mp)}\exp\left(-i\Delta\tilde{E}_{j1}^{(\mp)}L\right)
\tilde{U}_{\alpha j}^{(\mp)^\ast}\right|^2
\nonumber\\
&{\ }&\hspace*{-10mm}
=\left|\delta_{\alpha\beta}
-\sum_{j=1}^3
\tilde{U}_{\beta j}^{(\mp)}\left\{1-\exp\left(-i\Delta\tilde{E}_{j1}^{(\mp)}L\right)
\right\}
\tilde{U}_{\alpha j}^{(\mp)^\ast}\right|^2
\nonumber\\
&{\ }&\hspace*{-10mm}
=\left|\delta_{\alpha\beta}
-2i\exp\left(-\frac{i}{2}\Delta\tilde{E}_{j1}^{(\mp)}L\right)
\sum_{j=2}^3
\tilde{U}_{\beta j}^{(\mp)}\tilde{U}_{\alpha j}^{(\mp)^\ast}
\sin\left(\frac{\Delta\tilde{E}_{j1}^{(\mp)}L}{2}\right)
\right|^2\,,
\label{probv}
\end{eqnarray}
where we have used the unitarity property
$\sum_{j=1}^3 \tilde{U}_{\beta j}^{(\mp)}
\tilde{U}_{\alpha j}^{(\mp)^\ast} =\delta_{\alpha\beta}$
in the third line
and we have defined
\begin{eqnarray}
  &{\ }&\hspace*{-66mm}
  \Delta \tilde{E}_{jk}^{(\mp)}\equiv\tilde{E}_j^{(\mp)}
  -\tilde{E}_k^{(\mp)}\,.
\nonumber
\end{eqnarray}
Thus we can obtain the analytic expression if
we get the bilinear form
$\tilde{U}_{\beta j}^{(\mp)}\tilde{U}_{\alpha j}^{(\mp)^\ast}$.

It was shown by Kimura, Takamura and
Yokomakura\,\cite{Kimura:2002hb,Kimura:2002wd}
that $\tilde{U}_{\beta j}^{(\mp)}\tilde{U}_{\alpha j}^{(\mp)^\ast}$ can
be expressed in terms of the known quantities
as long as the energy eigenvalue $\tilde{E}_j^{(\mp)}$ is known.
Their argument goes as follows.
If we consider the ($\alpha$, $\beta$)-component of
$n$-th power ($n=0,1,2$) of the neutrino part of
Eq.\,(\ref{sch1}), then we obtain
\begin{eqnarray}
  &{\ }&\hspace*{-16mm}
\delta_{\alpha\beta}=\left[\tilde{U}^{(-)}(\tilde{U}^{(-)})^{-1}\right]_{\alpha\beta}
=\sum_j\tilde{U}_{\alpha j}^{(-)}(\tilde{U}_{\beta j}^{(-)})^\ast
\label{const1}\\
&{\ }&\hspace*{-16mm}
\left[U{\cal E}U^{-1}+{\cal A}\right]_{\alpha\beta}
=\left[\tilde{U}^{(-)}\tilde{{\cal E}^{(-)}}(\tilde{U}^{(-)})^{-1}\right]_{\alpha\beta}
=\sum_j\tilde{E}_j^{(-)}\tilde{U}_{\alpha j}^{(-)}(\tilde{U}_{\beta j}^{(-)})^\ast
\label{const2}\\
&{\ }&\hspace*{-16mm}
\left[\left(U{\cal E}U^{-1}+{\cal A}\right)^2\right]_{\alpha\beta}
=\left[\tilde{U}^{(-)}\tilde{({\cal E}}^{(-)})^2\tilde{U}^{-1}\right]_{\alpha\beta}
=\sum_j(\tilde{E}^{(-)}_j)^2\tilde{U}_{\alpha j}^{(-)}(\tilde{U}_{\beta j}^{(-)})^\ast\,.
\label{const3}
\end{eqnarray}
Putting Eqs. (\ref{const1})--(\ref{const3}) together,
we have
\begin{eqnarray}
\left(\begin{array}{ccc}
1&1&1\cr
\tilde{E}_1^{(-)}&\tilde{E}_2^{(-)}&\tilde{E}_3^{(-)}\cr
(\tilde{E}_1^{(-)})^2&(\tilde{E}_2^{(-)})^2&(\tilde{E}_3^{(-)})^2
\end{array}\right)
\left(\begin{array}{c}
  \tilde{U}_{\beta 1}^{(-)}\tilde{U}_{\alpha 1}^{(-)\ast}
\cr\cr
  \tilde{U}_{\beta 2}^{(-)}\tilde{U}_{\alpha 2}^{(-)\ast}
\cr\cr
  \tilde{U}_{\beta 3}^{(-)}\tilde{U}_{\alpha 3}^{(-)\ast}
\end{array}\right)
=\left(\begin{array}{r}
\delta_{\alpha\beta}\cr
\left[U{\cal E}U^{-1}+{\cal A}\right]_{\alpha\beta}\cr
\left[\left(U{\cal E}U^{-1}+{\cal A}\right)^2\right]_{\alpha\beta}
\end{array}\right),
\nonumber
\end{eqnarray}
which can be easily solved by inverting the
Vandermonde matrix:
\begin{eqnarray}
&{\ }&\hspace*{-12mm}
\left(\begin{array}{c}
  \tilde{U}_{\beta 1}^{(-)}\tilde{U}_{\alpha 1}^{(-)\ast}
\cr\cr
  \tilde{U}_{\beta 2}^{(-)}\tilde{U}_{\alpha 2}^{(-)\ast}
\cr\cr
  \tilde{U}_{\beta 3}^{(-)}\tilde{U}_{\alpha 3}^{(-)\ast}
\end{array}\right)
=\left(\begin{array}{ccc}
\displaystyle
\frac{{\ }1}{\Delta \tilde{E}_{21}^{(-)} \Delta \tilde{E}_{31}^{(-)}}
(\tilde{E}_2^{(-)}\tilde{E}_3^{(-)}, & -(\tilde{E}_2^{(-)}+\tilde{E}_3^{(-)}),&
1)\cr
\displaystyle
\frac{-1}{\Delta \tilde{E}_{21}^{(-)} \Delta \tilde{E}_{32}^{(-)}}
(\tilde{E}_3^{(-)}\tilde{E}_1^{(-)}, & -(\tilde{E}_3^{(-)}+\tilde{E}_1^{(-)}),&1)\cr
\displaystyle
\frac{{\ }1}{\Delta \tilde{E}_{31}^{(-)} \Delta \tilde{E}_{32}^{(-)}}
(\tilde{E}_1^{(-)}\tilde{E}_2^{(-)}, & -(\tilde{E}_1^{(-)}+\tilde{E}_2^{(-)}),&1)\cr
\end{array}\right)
\left(\begin{array}{r}
\delta_{\alpha\beta}\cr\cr
\left[U{\cal E}U^{-1}+{\cal A}\right]_{\alpha\beta}\cr\cr
\left[\left(U{\cal E}U^{-1}+{\cal A}\right)^2\right]_{\alpha\beta}
\end{array}\right)\,.
\nonumber\\
\label{solx}
\end{eqnarray}
The expression $\tilde{U}_{\beta j}^{(+)}\tilde{U}_{\alpha j}^{(+)\ast}$
for antineutrinos can be obtained in the same manner.
Eq.\,(\ref{probv}) together with (\ref{solx})
is exact in the case with constant density of matter,
as long as we know the energy eigenvalues $\tilde{E}_j^{(\mp)}$ exactly.

\section{The oscillation probabilities
for $|\Delta m_{31}^2/2E| \gg A \sim \Delta m_{21}^2/2E$}
\label{appendixb}

In the case of low energy accelerator neutrinos,
we have $|\Delta E_{31}|\equiv|\Delta m_{31}^2/2E|
\gg A \simeq \Delta E_{21} \equiv\Delta m_{21}^2/2E$,
so we keep $\Delta E_{31}$ and treat $A$ and $\Delta E_{21}$
as perturbation, keeping only terms of first order in $A$ and $\Delta E_{21}$.

The eigenvalues can be obtained from the eigenequation
\begin{eqnarray}
  &{\ }&\hspace*{-36mm}0= \det(t \mbox{\bf 1}-M^{(\mp)})
  \nonumber\\
&{\ }&\hspace*{-32mm}=t^3
-t^2 \mbox{\rm Tr}[M^{(\mp)}]
+\frac{t}{2} \left\{(\mbox{\rm Tr}[M^{(\mp)}])^2-\mbox{\rm Tr}[(M^{(\mp)})^2]\right\}
\nonumber\\
&{\ }&\hspace*{-28mm}
-\frac{1}{6} \left\{(\mbox{\rm Tr}[M^{(\mp)}])^3+2 \mbox{\rm Tr}[(M^{(\mp)})^3]
-3 \mbox{\rm Tr}[M^{(\mp)}] \mbox{\rm Tr}[(M^{(\mp)})^2]\right\}\,.
\label{eq1}
\end{eqnarray}
Here the matrix can be expressed as
\begin{eqnarray}
  &{\ }&\hspace*{-16mm}
  M^{(\mp)}\equiv
\left\{\begin{array}{l}
U{\cal E}U^{-1}{\cal A}
  =\Delta E_{31} U\eta_3 U^{-1}
  +\Delta E_{21} U\eta_2 U^{-1}+{\cal A}\\
U^\ast{\cal E}(U^\ast)^{-1}-{\cal A}
  =\Delta E_{31} U^\ast\eta_3 (U^\ast)^{-1}
  +\Delta E_{21} U^\ast\eta_2 (U^\ast)^{-1}-{\cal A}
  \end{array}\right\}
  \label{m}
\end{eqnarray}
where we have defined
\begin{eqnarray}
  &{\ }&\hspace*{-86mm}
  \eta_3 \equiv {\mbox{\rm diag}}\left(0,0,1\right)
  \nonumber\\
  &{\ }&\hspace*{-86mm}
  \eta_2 \equiv {\mbox{\rm diag}}\left(0,1,0\right)\,.
  \nonumber
\end{eqnarray}
In the last equation in Eq.\,(\ref{m}),
the first term is large while the
second and third terms are of order
$\Delta m^2_{21}/|\Delta m^2_{31}|\simeq 1/30$.
Applying a perturbation method
with respect to $\Delta m^2_{21}/|\Delta m^2_{31}|$, we obtain the
following eigenvalues for Eq.\,(\ref{eq1}) to the
leading order in $\Delta m^2_{21}/|\Delta m^2_{31}|$:
\begin{eqnarray}
  &{\ }&\hspace*{-50mm}
\left\{\begin{array}{l}
\tilde{E}_1^{(\mp)} =
\frac{1}{2}\left(\Delta E_{21} \pm \mbox{\rm Tr}[{\cal A}']
\mp\mbox{\rm Tr}[\eta_3 {\cal A}']
-\Delta \tilde{E}_{21}^{(\mp)}\right)
\\
\tilde{E}_2^{(\mp)} =
\frac{1}{2}\left(\Delta E_{21} \pm \mbox{\rm Tr}[{\cal A}']
\mp\mbox{\rm Tr}[\eta_3 {\cal A}']
+\Delta \tilde{E}_{21}^{(\mp)}\right)
\\
\tilde{E}_3^{(\mp)} = \Delta E_{31} \pm \mbox{\rm Tr}[\eta_3 {\cal A}']\,,
\end{array}\right\}
\label{e1}
\end{eqnarray}
where
$\Delta \tilde{E}_{21}$
is defined by by Eq.\,(\ref{tdele21}).
In obtaining Eq.\,(\ref{e1}),
we have used the properties of hermitian matrices ${\cal A}$
and ${\cal A}'\equiv\tilde{R}_{13}^{-1}\,R_{23}^{-1}\,{\cal A}\,
  R_{23}\, \tilde{R}_{13}$ which is defined in Eq.\,(\ref{ap1}):
\begin{eqnarray}
  &{\ }&\hspace*{-45mm}
  \mbox{\rm Tr}[{\cal A}]
  =\mbox{\rm Tr}[R_{23}\,\tilde{R}_{13}\,{\cal A}'
    \,\tilde{R}_{13}^{-1}\,R_{23}^{-1}]
  =\mbox{\rm Tr}[{\cal A}']
  \nonumber\\
  &{\ }&\hspace*{-45mm}
  \mbox{\rm Tr}[U \eta_3 U^{-1}{\cal A}]
  =\mbox{\rm Tr}[\eta_3 R_{12}^{-1}\,{\cal A}'
    \,R_{12}]
  =\mbox{\rm Tr}[\eta_3 {\cal A}']
  \nonumber\\
  &{\ }&\hspace*{-45mm}
  \mbox{\rm Tr}[{\cal A}^\ast]=\mbox{\rm Tr}[({\cal A}^\ast)^T]
  =\mbox{\rm Tr}[{\cal A}]
  \nonumber\\
  &{\ }&\hspace*{-45mm}
  \mbox{\rm Tr}[({\cal A}')^\ast]=\mbox{\rm Tr}[\{({\cal A}')^\ast\}^T]
  =\mbox{\rm Tr}[{\cal A}']
\nonumber
\end{eqnarray}
$\tilde{E}_3^{(\mp)}$ in Eq.\,(\ref{e1})
is given to the next leading order in
$\Delta m^2_{21}/|\Delta m^2_{31}|$
because it is necessary to obtain
$\tilde{U}_{e2}^{(\mp)}\tilde{U}_{\mu 2}^{(\mp)\ast }$ later.

For simplicity, we obtain the bilinear form
$\tilde{U}_{\beta j}^{(-)}\tilde{U}_{\alpha j}^{(-)\ast}$
for neutrinos only in the following.  The one
$\tilde{U}_{\beta j}^{(+)}\tilde{U}_{\alpha j}^{(+)\ast}$
for antineutrinos can be read off from the expression
$\tilde{U}_{\beta j}^{(-)}\tilde{U}_{\alpha j}^{(-)\ast}$.
Let us introduce the notation
\begin{eqnarray}
&{\ }&\hspace{-80mm}
Y_j^{\alpha\beta}\equiv
\left[\left(
U{\cal E}U^{-1}+{\cal A}\right)^{j-1}\right]_{\alpha\beta}
\nonumber
\end{eqnarray}
To perform perturbation calculations,
it is convenient to rescale $\Delta E_{21}\to\epsilon\Delta E_{21}$
and ${\cal A}_{\alpha\beta}\to\epsilon{\cal A}_{\alpha\beta}$.
Then we have
\begin{eqnarray}
&{\ }&\hspace{-20mm}
Y_1^{e\mu}=0
\nonumber\\
&{\ }&\hspace{-20mm}
Y_2^{e\mu}=\left(U{\cal E}U^{-1}+{\cal A}\right)_{e\mu}
\nonumber\\
&{\ }&\hspace{-12mm}
=\Delta E_{31} U_{e3}U_{\mu 3}^\ast + \epsilon\left(
\Delta E_{21} U_{e2}U_{\mu 2}^\ast+{\cal A}_{e\mu}\right)
\nonumber\\
&{\ }&\hspace{-20mm}
Y_3^{e\mu}=\left[\left(
U{\cal E}U^{-1}+{\cal A}\right)^2\right]_{e\mu}
\nonumber\\
&{\ }&\hspace{-12mm}
=\Delta E_{31}^2 U_{e3}U_{\mu 3}^\ast
+ \epsilon \Delta E_{31}
\{U\eta_3 U^{-1}, {\cal A}\}_{e\mu}
\nonumber\\
&{\ }&\hspace{-8mm}
+ \epsilon^2 \left[
\left(\Delta E_{21}\right)^2 U_{e2}U_{\mu 2}^\ast
+\Delta E_{21} U_{e2}U_{\mu 2}^\ast
\{U\eta_2 U^{-1}, {\cal A}\}_{e\mu}
+\left({\cal A}^2\right)_{e\mu}\right]
\nonumber
\end{eqnarray}
With these quantities, we have Eq.\,(\ref{txme3}) to the leading order in
${\cal O}(\epsilon)$:
\begin{eqnarray}
&{\ }&\hspace{-46mm}
\tilde{U}_{e3}^{(-)}(\tilde{U}_{\mu 3}^{(-)})^\ast=
\frac{-(\tilde{E}_1^{(-)}+\tilde{E}_2^{(-)}) Y_2^{e\mu}+Y_3^{e\mu}}
{(\tilde{E}_3^{(-)}-\tilde{E}_1^{(-)}) (\tilde{E}_3^{(-)}-\tilde{E}_2^{(-)})}
     \simeq
 U_{e3}U_{\mu 3}^\ast\,.
\nonumber
\end{eqnarray}
In the case of antineutrinos, we have to replace
$U_{\alpha j}$ by $U_{\alpha j}^\ast$, and we have Eq.\,(\ref{txmeb3})
\begin{eqnarray}
&{\ }&\hspace{-90mm}
\tilde{U}_{e3}^{(+)}(\tilde{U}_{\mu 3}^{(+)})^\ast\simeq
 U_{e3}^\ast U_{\mu 3}\,.
\nonumber
\end{eqnarray}
For $\tilde{U}_{e2}\tilde{U}_{\mu 2}^\ast$,
we obtain Eq.\,(\ref{txme2}) to the leading order in
${\cal O}(\epsilon)$:
\begin{eqnarray}
&{\ }&\hspace{-8mm}
\tilde{U}_{e2}^{(-)}\tilde{U}_{\mu 2}^{(-)\ast}=
\frac{(\tilde{E}_3^{(-)}+\tilde{E}_1^{(-)}) Y_2^{e\mu}-Y_3^{e\mu}}
{(\tilde{E}_3^{(-)}-\tilde{E}_2^{(-)})(\tilde{E}_2^{(-)}-\tilde{E}_1^{(-)})}
\nonumber\\
&{\ }&\hspace{5mm}
\simeq
\frac{1}{\Delta \tilde{E}_{21}^{(-)}}\left\{
\Delta E_{21}U_{e2}U_{\mu 2}^\ast
+\left(
\Delta E_{21} -\Delta \tilde{E}_{21}^{(-)}\right)
\frac{U_{e3}U_{\mu 3}^\ast}{2}\right\}
\nonumber\\
&{\ }&\hspace{7mm}
+\frac{1}{\Delta \tilde{E}_{21}^{(-)}}\left[
{\cal A}_{e\mu}
-\{U\eta_3 U^{-1}, {\cal A}\}_{e\mu}
+\left(
\mbox{\rm Tr}[{\cal A}]
+\mbox{\rm Tr}[U \eta_3 U^{-1}{\cal A}]
\right)\frac{U_{e3}U_{\mu 3}^\ast}{2}\right]
\nonumber\\
&{\ }&\hspace{5mm}
\simeq
\frac{1}{\Delta \tilde{E}_{21}^{(-)}}\left[
\Delta E_{21}U_{e2}U_{\mu 2}^\ast
+\left\{
\Delta E_{21} -\Delta \tilde{E}_{21}^{(-)}
+A\left(2\epsilon_D-c_{13}^2\right)
\right\}
\frac{U_{e3}U_{\mu 3}^\ast}{2}
+A\epsilon_{N}U_{\tau 3}\right]
\nonumber
\end{eqnarray}
\noindent
$\epsilon_I$, $\epsilon_D$ and $\epsilon_N$
are defined in Eqs.\,(\ref{epsiloni}), (\ref{epsilond})
and (\ref{epsilonn}).
In the case of antineutrinos, we have to replace
$U_{\alpha j}$ by $U_{\alpha j}^\ast$ and
${\cal A}$ by $-{\cal A}$, and we have Eq.\,(\ref{txmeb2})
\begin{eqnarray}
&{\ }&\hspace{-8mm}
\tilde{U}_{e2}^{(+)}(\tilde{U}_{\mu 2}^{(+)})^\ast
\simeq
\frac{1}{\Delta \tilde{E}_{21}^{(-)}}\left[
  \Delta E_{21}U_{e2}^\ast U_{\mu 2}
+\left\{
\Delta E_{21} -\Delta \tilde{E}_{21}^{(-)}
-A\left(2\epsilon_D-c_{13}^2\right)
\right\}
\frac{U_{e3}^\ast U_{\mu 3}}{2}
-A\epsilon_{N}U_{\tau 3}\right]\,.
\nonumber
\end{eqnarray}
\noindent
As for the disappearance channel,
on the other hand, we have the following:
\begin{eqnarray}
&{\ }&\hspace{-20mm}
Y_1^{\mu\mu}=1
\nonumber\\
&{\ }&\hspace{-20mm}
Y_2^{\mu\mu}=\left(U{\cal E}U^{-1}+{\cal A}\right)_{\mu\mu}
\nonumber\\
&{\ }&\hspace{-12mm}
=\Delta E_{31} |U_{\mu 3}|^2 + \epsilon\left(
\Delta E_{21} |U_{\mu 2}|^2+{\cal A}_{\mu\mu}\right)
\nonumber\\
&{\ }&\hspace{-20mm}
Y_3^{\mu\mu}=\left\{\left(
U{\cal E}U^{-1}+{\cal A}\right)\right\}_{\mu\mu}
\nonumber\\
&{\ }&\hspace{-12mm}
=\Delta E_{31}^2 |U_{\mu 3}|^2
+ \epsilon \Delta E_{31}
\{U\eta_3 U^{-1}, {\cal A}\}_{\mu\mu}
\nonumber\\
&{\ }&\hspace{-8mm}
+ \epsilon^2 \left[
\left(\Delta E_{21}\right)^2 |U_{\mu 2}|^2
+\Delta E_{21} |U_{\mu 2}|^2
\{U\eta_2 U^{-1}, {\cal A}\}_{\mu\mu}
+\left({\cal A}^2\right)_{\mu\mu}\right]
\nonumber
\end{eqnarray}
The bilinear form $|\tilde{U}_{\mu 3}^{(-)}|^2$
is thus given by Eq.\,(\ref{txmm3}):
\begin{eqnarray}
&{\ }&\hspace{-25mm}
|\tilde{U}_{\mu 3}^{(-)}|^2=
\frac{-(\tilde{E}_1^{(-)}+\tilde{E}_2^{(-)}) Y_2^{\mu\mu}+Y_3^{\mu\mu}}
{(\tilde{E}_3^{(-)}-\tilde{E}_1^{(-)}) (\tilde{E}_3^{(-)}-\tilde{E}_2^{(-)})}
     \simeq
       |U_{\mu 3}|^2\,.
\nonumber
\end{eqnarray}
$|\tilde{U}_{\mu 3}^{(+)}|^2$ is also equal to the one in vacuum and
therefore is given by Eq.\,(\ref{txmm3}).
$|\tilde{U}_{\mu 2}^{(-)}|^2$ is given by Eq.\,(\ref{txmm2}):
\begin{eqnarray}
&{\ }&\hspace*{-23mm}
|\tilde{U}_{\mu 2}^{(-)}|^2=
-\frac{\tilde{E}_3^{(-)} \tilde{E}_1^{(-)}-(\tilde{E}_3^{(-)}+\tilde{E}_1^{(-)}) Y_2^{\mu\mu}+Y_3^{\mu\mu}}
{(\tilde{E}_3^{(-)}-\tilde{E}_1^{(-)})(\tilde{E}_2^{(-)}-\tilde{E}_1^{(-)})}
\nonumber\\
&{\ }&\hspace{-12mm}
\simeq
\frac{1}{\Delta \tilde{E}_{21}^{(-)}}\left\{
\Delta E_{21} |U_{\mu 2}|^2+
 \left(\Delta E_{21}- \Delta \tilde{E}_{21}^{(-)}\right)
 \frac{|U_{\mu 3}|^2}{2}\right\}
\nonumber\\
&{\ }&\hspace{-10mm}
+\frac{1}{\Delta \tilde{E}_{21}^{(-)}}\left[
{\cal A}_{\mu\mu}
-\{U\eta_3 U^{-1}, {\cal A}\}_{\mu\mu}
+\left(
\mbox{\rm Tr}[{\cal A}]
+\mbox{\rm Tr}[U \eta_3 U^{-1}{\cal A}]
\right)\frac{|U_{\mu 3}|^2}{2}\right]
\nonumber\\
&{\ }&\hspace{-12mm}
  \simeq\frac{1}{\Delta \tilde{E}_{21}^{(-)}}\left[
 \Delta E_{21} |U_{\mu 2}|^2+
 \left(\Delta E_{21}- \Delta \tilde{E}_{21}^{(-)}\right)
 \frac{|U_{\mu 3}|^2}{2}\right.
+ A\left(1+c_{23}^2\right)
\nonumber\\
&{\ }&\hspace{10mm}
\left. +2 A\,\left( \epsilon_I- \epsilon_D\,c_{23}^2\right)
\right]
\nonumber
\end{eqnarray}
In the case of antineutrinos, we have to replace
${\cal A}$ by $-{\cal A}$, so we get 
\begin{eqnarray}
&{\ }&\hspace*{-23mm}
  |\tilde{U}_{\mu 2}^{(+)}|^2
  \simeq\frac{1}{\Delta \tilde{E}_{21}^{(+)}}\left[
 \Delta E_{21} |U_{\mu 2}|^2+
 \left(\Delta E_{21}- \Delta \tilde{E}_{21}^{(+)}\right)
 \frac{|U_{\mu 3}|^2}{2}\right.
- A\left(1+c_{23}^2\right)
\nonumber\\
&{\ }&\hspace{10mm}
\left. -2 A\,\left( \epsilon_I- \epsilon_D\,c_{23}^2\right)
\right]
\nonumber
\end{eqnarray}
as in Eq.\,(\ref{txmm2}).

\section{Derivation of $f^{(\mp)}$ in (\ref{fmp}) and $g^{(\mp)}$ in (\ref{gmp})}
\label{appendixc}
From Eq.\,(\ref{pmm}), using Eqs.\,(\ref{txmm3}), (\ref{txmm2}),
and (\ref{tdele21v2})--(\ref{tdele21v4}),
we have
\begin{eqnarray}
&{\ }&\hspace*{-6mm}
  \left\{\begin{array}{l}
  P(\nu_{\mu}\to\nu_{\mu})\\
P(\bar{\nu}_{\mu}\to \bar{\nu}_\mu)  
  \end{array}\right\}
  \nonumber\\
&{\ }&\hspace*{-10mm}
\simeq \left|
1-2i
e^{-i\Delta E_{31} L/2}
|{U}_{\mu 3}|^2
\sin\left(\frac{\Delta E_{31} L}{2}\right)
-2i\exp\left(-i\frac{L}{2}
  \left.\Delta
  \tilde E_{21}^{(\mp)}\right|_{\mbox{\tiny\rm std}}\right)
e^{-i\delta\Delta \tilde E_{21}^{(\mp)} L/2}
|\tilde{U}_{\mu 2}^{(\mp)}|^2
\frac{\Delta \tilde E_{21}^{(\mp)} L}{2}
\right|^2
\nonumber\\
&{\ }&\hspace*{-10mm}
= 4\left|
\exp\left(i\frac{L}{2}\left.\Delta \tilde E_{21}^{(\mp)}\right|_{\mbox{\tiny\rm std}} \right)
\left\{\frac{i}{2}
+e^{-i\Delta E_{31} L/2}
|{U}_{\mu 3}|^2
\sin\left(\frac{\Delta E_{31} L}{2}\right)\right\}
+e^{-i\delta\Delta \tilde E_{21}^{(\mp)} L/2}
|\tilde{U}_{\mu 2}^{(\mp)}|^2
\frac{\Delta \tilde E_{21}^{(\mp)} L}{2}
\right|^2
\nonumber\\
&{\ }&\hspace*{-10mm}
\simeq 4\left|
\exp\left(i\frac{L}{2}\left.\Delta \tilde E_{21}^{(\mp)}\right|_{\mbox{\tiny\rm std}} \right)
\left\{\frac{i}{2}
+e^{-i\Delta E_{31} L/2}
|{U}_{\mu 3}|^2
\sin\left(\frac{\Delta E_{31} L}{2}\right)\right\}\right.
\nonumber\\
&{\ }&\hspace*{-5mm}
+\left(1-i\frac{L}{2}\delta\Delta \tilde E_{21}^{(\mp)}\right)
\left[
   \frac{\Delta E_{21} L}{2} |U_{\mu 2}|^2+
 \left(\frac{\Delta E_{21}L}{2}
 - \frac{\left.\Delta \tilde E_{21}^{(\mp)}\right|_{\mbox{\tiny\rm std}} L
   +\delta\Delta \tilde{E}_{21}^{(\mp)}L}{2}\right)
 \frac{|U_{\mu 3}|^2}{2}
 \right.
\nonumber\\
&{\ }&\hspace{35mm}
\left.\pm \frac{AL}{2}\left(
1+c_{23}^2
+2 \, \epsilon_I
+2 \, \epsilon_D
\left.c_{23}^2 \right)\right]\right|^2
\nonumber\\
&{\ }&\hspace*{-10mm}
\simeq 4\left|
\exp\left(i\frac{L}{2}\left.\Delta \tilde E_{21}^{(\mp)}\right|_{\mbox{\tiny\rm std}} \right)
\left\{\frac{i}{2}
+e^{-i\Delta E_{31} L/2}
|{U}_{\mu 3}|^2
\sin\left(\frac{\Delta E_{31} L}{2}\right)\right\}\right.
\nonumber\\
&{\ }&\hspace*{-5mm}
+   \frac{\Delta E_{21} L}{2} |U_{\mu 2}|^2+
 \left(\frac{L}{2}\Delta E_{21}
 - \frac{L}{2}\left.\Delta \tilde E_{21}^{(\mp)}\right|_{\mbox{\tiny\rm std}} \right)
 \frac{|U_{\mu 3}|^2}{2}\pm \frac{AL}{2}\left(
1+c_{23}^2\right)
\nonumber\\
&{\ }&\hspace{-5mm}
-i\frac{L}{2}\delta\Delta \tilde E_{21}^{(\mp)}
\left\{
   \frac{\Delta E_{21} L}{2} |U_{\mu 2}|^2+
 \left(\frac{L}{2}\Delta E_{21}
 - \frac{L}{2}\left.\Delta \tilde E_{21}^{(\mp)}\right|_{\mbox{\tiny\rm std}}
 -i\right)
 \frac{|U_{\mu 3}|^2}{2}
 \right\}
\nonumber\\
&{\ }&\hspace{-5mm}
\pm AL\left(
 \, \epsilon_I
+ \, \epsilon_D
\left.c_{23}^2 \right)\right|^2\,,
\nonumber
\end{eqnarray}
where Eq.\,(\ref{solarterm2}) was used in the third step above.
Here introducing the notation
\begin{eqnarray}
&{\ }&\hspace*{-46mm}
{\cal F}\equiv   \frac{\Delta E_{21} L}{2} |U_{\mu 2}|^2+
 \left(\frac{L}{2}\Delta E_{21}
 - \frac{L}{2}\left.\Delta \tilde E_{21}^{(\mp)}\right|_{\mbox{\tiny\rm std}}
 -i\right)
 \frac{|U_{\mu 3}|^2}{2}
\nonumber
\end{eqnarray}
as in Eq.\,(\ref{f}), we get
\begin{eqnarray}
&{\ }&\hspace*{-6mm}
  \left\{\begin{array}{l}
  P(\nu_{\mu}\to\nu_{\mu})\\
P(\bar{\nu}_{\mu}\to \bar{\nu}_\mu)  
  \end{array}\right\}
  \nonumber\\
&{\ }&\hspace*{-10mm}
\simeq 4\left|
\exp\left(i\frac{L}{2}\left.\Delta \tilde E_{21}^{(\mp)}\right|_{\mbox{\tiny\rm std}} \right)
\left\{\frac{i}{2}
+e^{-i\Delta E_{31} L/2}
|{U}_{\mu 3}|^2
\sin\left(\frac{\Delta E_{31} L}{2}\right)\right\}\right.
\nonumber\\
&{\ }&\hspace*{-5mm}
+   \frac{\Delta E_{21} L}{2} |U_{\mu 2}|^2+
 \left(\frac{L}{2}\Delta E_{21}
 - \frac{L}{2}\left.\Delta \tilde E_{21}^{(\mp)}\right|_{\mbox{\tiny\rm std}} \right)
 \frac{|U_{\mu 3}|^2}{2}\pm \frac{AL}{2}\left(
1+c_{23}^2\right)
\nonumber\\
&{\ }&\hspace{-5mm}
\mp i\frac{AL}{\left.\Delta \tilde{E}_{21}^{(\mp)}\right|_{\mbox{\footnotesize\rm std}}}
\left\{\epsilon_D\left(\Delta E_{21}\cos2\theta_{12}\mp Ac_{13}^2 \right)
  + \mbox{\rm Re}(\epsilon_N) \Delta E_{21}\sin2\theta_{12}\right\}\,{\cal F}
\nonumber\\
&{\ }&\hspace{-5mm}
\pm AL\left(
 \, \epsilon_I
+ \, \epsilon_D
\left.c_{23}^2 \right)\right|^2
\nonumber\\
&{\ }&\hspace*{-10mm}
= 4\left|
\exp\left(i\frac{L}{2}\left.\Delta \tilde E_{21}^{(\mp)}\right|_{\mbox{\tiny\rm std}} \right)
\left\{\frac{i}{2}
+e^{-i\Delta E_{31} L/2}
|{U}_{\mu 3}|^2
\sin\left(\frac{\Delta E_{31} L}{2}\right)\right\}\right.
\nonumber\\
&{\ }&\hspace*{-5mm}
+   \frac{\Delta E_{21} L}{2} |U_{\mu 2}|^2+
 \left(\frac{L}{2}\Delta E_{21}
 - \frac{L}{2}\left.\Delta \tilde E_{21}^{(\mp)}\right|_{\mbox{\tiny\rm std}} \right)
 \frac{|U_{\mu 3}|^2}{2}\pm \frac{AL}{2}\left(
1+c_{23}^2\right)
\nonumber\\
&{\ }&\hspace{-5mm}
\left.\pm AL\left[
 \, \epsilon_I
 + \, \epsilon_D
 \left(
 c_{23}^2-i\frac{\Delta E_{21}\cos2\theta_{12}\mp Ac_{13}^2}
 {\left.\Delta \tilde{E}_{21}^{(\mp)}\right|_{\mbox{\footnotesize\rm std}}}{\cal F}
 \right)
 -i\mbox{\rm Re}(\epsilon_N)
 \frac{\Delta E_{21}\sin2\theta_{12}}
 {\left.\Delta \tilde{E}_{21}^{(\mp)}\right|_{\mbox{\footnotesize\rm std}}}{\cal F}
  \right]\right|^2
\nonumber\\
\label{pmmv3}
\end{eqnarray}
Thus we get the expressions (\ref{fmp}) for $f^{(\mp)}$
and (\ref{gmp}) for $g^{(\mp)}$.

\end{document}